\documentclass[twocolumn, twocolappendix]{aastex631}

\usepackage{mathtools}
\usepackage{bm}
\usepackage{color}

\graphicspath{{./}{figures/}}

\begin{document}


\title{Connecting a Magnetized Disk to a Convective Low-mass Protostar:\\
A Global Three-dimensional Model of Boundary Layer Accretion}

\correspondingauthor{Shinsuke Takasao}
\email{shinsuke.takasao.astro@gmail.com}

\author[0000-0003-3882-3945]{Shinsuke Takasao}
\affiliation{Department of Earth and Space Science, Graduate School of Science, Osaka University, Toyonaka, Osaka 560-0043, Japan}

\author[0000-0003-3127-5982]{Takashi Hosokawa}
\affiliation{Department of Physics, Graduate School of Science, Kyoto University, Sakyo, Kyoto 606-8502, Japan}

\author[0000-0001-8105-8113]{Kengo Tomida}
\affiliation{Astronomical Institute, Tohoku University, Aoba, Sendai, Miyagi 980-0578, Japan}

\author[0000-0002-2707-7548]{Kazunari Iwasaki}
\affiliation{Center for Computational Astrophysics, National Astronomical Observatory of Japan, Mitaka, Tokyo 181-8588, Japan}



\begin{abstract}
In the early stages of star formation, boundary layer accretion, where protostars accrete material from disks extending down to their surfaces, plays a crucial role. Understanding how a magneto-rotational-instability (MRI)-active disk connects to a protostar's surface remains a significant challenge. To investigate the mechanisms of mass and angular momentum transfer, we develop a global, three-dimensional magnetohydrodynamic model of boundary layer accretion around a magnetized, convective low-mass protostar. 
Our results reveal that angular momentum transport mechanisms transition significantly from the outer MRI-active disk to the protostellar surface. Various mechanisms—MRI, spiral shocks, coronal accretion, jets, and disk winds—contribute to angular momentum transfer, resulting in three distinct disk structures: (1) the MRI-active disk, (2) the transition layer, and (3) the boundary layer. 
The simulated protostar is strongly magnetized due to the accumulation of the disk fields, wrapping by disk toroidal fields, and stellar dynamo activity. Magnetic concentrations analogous to starspots form on the protostar and interact with the rotating disk gas to generate spiral shocks. These shocks play a key role in driving accretion. These findings demonstrate the necessity of global MHD models for a comprehensive understanding of angular momentum transport.
Additionally, we identify explosive events triggered by magnetic reconnection in both the protostar and the disk atmosphere. We also find decretion flows in the disk midplane, which may be important for the radial transport of refractory materials, such as Calcium-Aluminium-rich Inclusions (CAIs) precursor gas, to the outer disk.
\end{abstract}



\section{Introduction} \label{sec:intro}

Mass accretion from a disk onto a protostar is a key process that regulates protostellar evolution and ultimately determines the star's final fate. The accretion structure near a protostar is not yet fully understood, but two modes have been widely discussed. If the protostar develops a magnetosphere that truncates the inner disk, magnetospheric accretion occurs \citep{Hartmann2016ARA&A}. In the absence of a magnetosphere, the disk extends down to the protostellar surface, which rotates more slowly than the Keplerian velocity. This results in the formation of a narrow layer where the angular velocity of the accreting gas adjusts to the protostar's rotation speed. The interface layer is called a boundary layer, and this accretion mode is known as boundary layer accretion \citep{Lynden-Bell1974MNRAS,Popham1993ApJ}.
Boundary layer accretion is expected to occur in various contexts, including accreting neutron stars \citep{Popham2001ApJ}, protoplanet formation \citep{Dong2021ApJ}, and formation of the Moon by a Giant Impact \citep{Mullen2020ApJ}.

In star formation, boundary layer accretion is particularly relevant for young protostars and rapid accretors, where stellar magnetospheres are less likely to develop. The boundary layer significantly impacts observable properties, potentially emitting up to half of the total accretion luminosity \citep{Lynden-Bell1974MNRAS}. 
Boundary layer accretion is the primary focus of this study.

The transfer of angular momentum in the boundary layer remains poorly understood. The magnetorotational instability (MRI) is believed to be the primary mechanism driving accretion in the well-ionized inner disk \citep[e.g.,][]{Gammie1996ApJ}. However, MRI is linearly stable in the boundary layer, where the angular velocity increases with radius.

As a result, hydrodynamic mechanisms for generating effective viscosity have garnered attention. Turbulence induced by the Kelvin-Helmholtz instability is unlikely, at least on stellar scales, due to the supersonic rotation of the disk gas. Instead, \citet{Belyaev2013ApJ} proposed that acoustic waves generated by shear-acoustic instabilities could carry away angular momentum and drive accretion \citep[see also][]{Belyaev2012ApJ}. Other hydrodynamic simulations also suggest the importance of non-local transport by acoustic waves \citep[e.g.,][]{Hertfelder2015A&A,Coleman2022MNRAS}. 

However, vertically unstratified 3D MHD simulations by \citet{Belyaev2018MNRAS} found that these acoustic waves transfer only a small fraction of the angular momentum. Consequently, angular momentum accumulates in the boundary layer, forming a rapidly rotating belt around the stellar equator \footnote{\citet{Pessah2012ApJ} also argued that net angular momentum transfer does not occur in the boundary layer under the shearing-box approximation. However, this approximation does not account for the star's location relative to the computational domain. In contrast, \citet{Belyaev2018MNRAS}'s results are based on 3D MHD simulations without the shearing-box assumption.}. 
Thus, local models do not exhibit steady-state accretion in the boundary layer.

Magnetic torques are expected to operate in the boundary layer, but the local simulations of \citet{Belyaev2018MNRAS} indicate that they are insufficient to achieve a steady state. It was previously speculated that magnetic fields in the boundary layer would be amplified by velocity shear, generating significant Maxwell stress \citep{Pringle1989MNRAS,Armitage2002MNRAS, Steinacker2002ApJ}. However, \citet{Belyaev2018MNRAS} found no clear evidence supporting this hypothesis. Instead, they showed that magnetic field amplification in the boundary layer primarily arises from the simple pile-up of magnetic flux with accreting gas.

While previous local models suggest that magnetic fields play only minor roles, they do not fully account for the vertical structure of the star-disk interface. When disks are threaded by poloidal magnetic fields, angular momentum can be extracted vertically via magnetically driven jets and disk winds. Additionally, magnetic braking drives faster accretion near the disk surfaces than in the equatorial plane, generating additional Maxwell torque \citep{Matsumoto1996ApJ,Zhu2018ApJ,Takasao2018ApJ,Jacquemin-Ide2021A&A}. This phenomenon, known as coronal accretion, is associated with efficient magnetic flux transport through a process termed the coronal mechanism \citep{Beckwith2009ApJ}.
These examples highlight the need for global models to comprehensively evaluate the roles of magnetic fields in boundary layer accretion.

In addition to the vertical structure, the impact of the star's magnetized surface has not been discussed.
Low-mass protostars possess convective envelopes \citep[e.g.,][]{Baraffe2010A&A}, which are expected to produce starspots similar to those observed on the Sun and other convective stars. The strength of the magnetic fields in these starspots is determined by the balance between magnetic pressure and gas pressure near the stellar surface \citep[e.g.,][]{Safier1999ApJ}. Consequently, the magnetic pressure from the starspots can shape the structure of the stellar surface by influencing the isosurfaces of total pressure (the sum of gas and magnetic pressures). 
When the rotating disk gas interacts with the stellar surface, stellar magnetic fields can dynamically influence its motion. 
However, most previous studies modeled the stellar surfaces as hydrostatic layers. \citet{Kley1996ApJ} studied boundary layer accretion of a convective star, but the model is unmagnetized.

A detailed understanding of boundary layer accretion is also crucial for uncovering the mechanisms of radial material transport in the early solar system. Refractory inclusions, such as Calcium-Aluminium-rich Inclusions (CAIs), are the oldest dated solids in the solar system \citep{Connelly2012Sci}. CAIs are believed to have formed in the inner, hot regions of the disk ($\gtrsim 1400$~K) \citep[e.g.,][]{Nittler2016ARA&A}. However, their frequent presence in carbonaceous chondrites found on Earth suggests radial transport to au scales \citep{Yang2012M&PS,Desch2018ApJS}. The formation site and transport processes remain unresolved.

The presence of the short-lived radionuclide $^{10}$Be in CAIs, considered to be produced by cosmic-ray spallation reactions, implies that CAIs formed in regions where cosmic rays generated by flares of the young Sun were abundant \citep{McKeegan2000Sci,Gounelle2013ApJ,Jacquet2019A&A}\footnote{Other hypotheses have also been proposed \citep[e.g.,][]{Desch2004ApJ,Banerjee2016NatCo}.}. If this is the case, understanding how CAIs were transported to outer disk regions is to be answered.

To investigate the mechanisms of boundary layer accretion during the early stages of star formation, we perform a global 3D MHD simulation of boundary layer accretion. Since protostars are expected to be convective and surrounded by strongly magnetized disks \citep{Machida2007ApJ,Hosokawa2009ApJ,Tsukamoto2015MNRAS,Tomida2015ApJ,Vaytet2018A&A}, our study focuses on accretion in this context. While advanced 2D and 3D simulations have examined accretion in the very early phases \citep{Ahmad2023A&A,Bhandare2020A&A,Kimura2023ApJ}, these models do not include magnetic fields. Previous 3D MHD simulations have adopted a hydrostatic stellar envelope \citep{Armitage2002MNRAS,Belyaev2018MNRAS}, neglecting the effects of stellar convection.
This study aims to explore how two key factors influence the structure of the boundary layer: global 3D MHD dynamics and stellar convection.

The remainder of this paper is organized as follows. Section~\ref{sec:method} details the numerical methods used in this study. We also introduce the setup of our global 3D MHD model. Section~\ref{sec:results} describes the accretion structure and the importance of magnetic fields. We also examines explosive phenomena associated with the release of magnetic energy. In Section~\ref{sec:discussion}, we discuss the origin of stellar magnetic fields, potential observational implications, and the limitations of our model. Finally, Section~\ref{sec:summary} summarizes our key findings.

\section{Method} \label{sec:method}
\subsection{Basic equations and numerical scheme}\label{subsec:basic_eq}
The basic equations are 3D resistive MHD equations:
\begin{align}
    \frac{\partial \rho}{\partial t}+\nabla\cdot(\rho \bm{v})=0,\\
    \frac{\partial \rho \bm{v}}{\partial t}+\nabla\cdot \left(\rho \bm{v}\bm{v}-\frac{\bm{B}\bm{B}}{4\pi}+\mathbf{P}^{\ast}\right)=\rho\bm{g},\\
    \frac{\partial \bm{B}}{\partial t}=-\nabla\times (\bm{v}\times \bm{B})-\nabla\times(\eta\nabla\times \bm{B}),\\
    \frac{\partial e}{\partial t}+\nabla\cdot \left[(e+P^{\ast})\bm{v}-\frac{\bm{B}}{4\pi}(\bm{B}\cdot\bm{v})\right] &= \nonumber\\
    \rho \bm{g}\cdot \bm{v} + \frac{1}{4\pi}\nabla\cdot(\bm{B}\times \eta \nabla\times \bm{B}) + \Lambda,
\end{align}
where $\rho$ is the density, $\bm{v}$ and $\bm{B}$ denote the velocity and magnetic field vectors, respectively. $\mathbf{P}^{\ast}$ is a diagonal tensor with the components $P^{\ast}=p+B^2/8\pi$, where $p$ represents the gas pressure. The total energy density $e$ is the sum of the internal, kinetic and magnetic energy densities.
We adopt the equation of state for an ideal gas with a specific heat ratio of $\gamma=5/3$. $\bm{g}$ is the gravitational acceleration vector. $\bm{J}=(c/4\pi)\nabla\times \bm{B}$ is the current density, and $\eta$ is the Ohm-type resistivity. 
The radiative cooling is modeled using a simplified cooling function, $\Lambda$, which will be described in Section~\ref{sec:cooling}.

We solve the basic equations in Cartesian coordinates $(x, y, z)$. For data analysis, we map data into both cylindrical and spherical coordinates, denoted by $(R, \varphi, z)$ and $(r, \theta, \varphi)$, respectively.

To solve these equations, we employ a modified version of Athena++ \citep{Stone2020ApJS}. The numerical time step can be significantly smaller than the orbital timescale of the disk, often by orders of magnitude. This reduction is due to high Alfv\'en speeds, which arise from the accumulation of poloidal fields in the low-density regions of the stellar polar areas.
To mitigate this issue, we adopt the Harten-Lax-van Leer Discontinuities (HLLD) approximate Riemann solver \citep{Miyoshi2005JCoPh} with the Boris correction \citep[Boris-HLLD;][]{Matsumoto2019ApJ}, which alleviates the stringent Courant-Friedrichs-Lewy (CFL) condition in highly magnetized regions by moderating the Alfv\'en speed.

In regions of strong magnetization, gas pressure can become negative if not properly managed. While a locally-isothermal equation of state can prevent this, it is unsuitable to study explosive phenomena due to magnetic reconnection (Section~\ref{subsec:explosive}). Alternatively, we employ the dual energy formalism \citep{Bryan1995CoPhC, Takasao2022ApJ, Zhong2024ApJ}, which solves the internal energy in a non-conservative form. This method is less susceptible to negative pressures and is applied specifically in high-risk areas.

The spatial reconstruction of primitive variables is performed using the piecewise parabolic method (PPM) \citep{Colella1984JCoPh}. We do not use the extremum-preserving limiters for PPM \citep{Colella2008JCoPh}, which are included in the public version of the code, to enhance numerical stability.

Time integration of the equations, excluding the diffusion terms, employs the strong-stability-preserving third-order Runge-Kutta method. 
In ideal MHD, magnetic reconnections are triggered by grid-scale dissipation which is completely numerical and not controllable. To mitigate this effect, we add explicit magnetic diffusivity so that magnetic reconnection sites can be consistently resolved. The diffusivity model is detailed in Section~\ref{sec:artificial_diff}. The diffusion terms are integrated using the first-order Runge-Kutta-Legendre super-timestepping scheme \citep{Meyer2014JCoPh}.

\subsection{Initial Condition}
\begin{figure*}
    \centering
    \includegraphics[width=2.0\columnwidth]{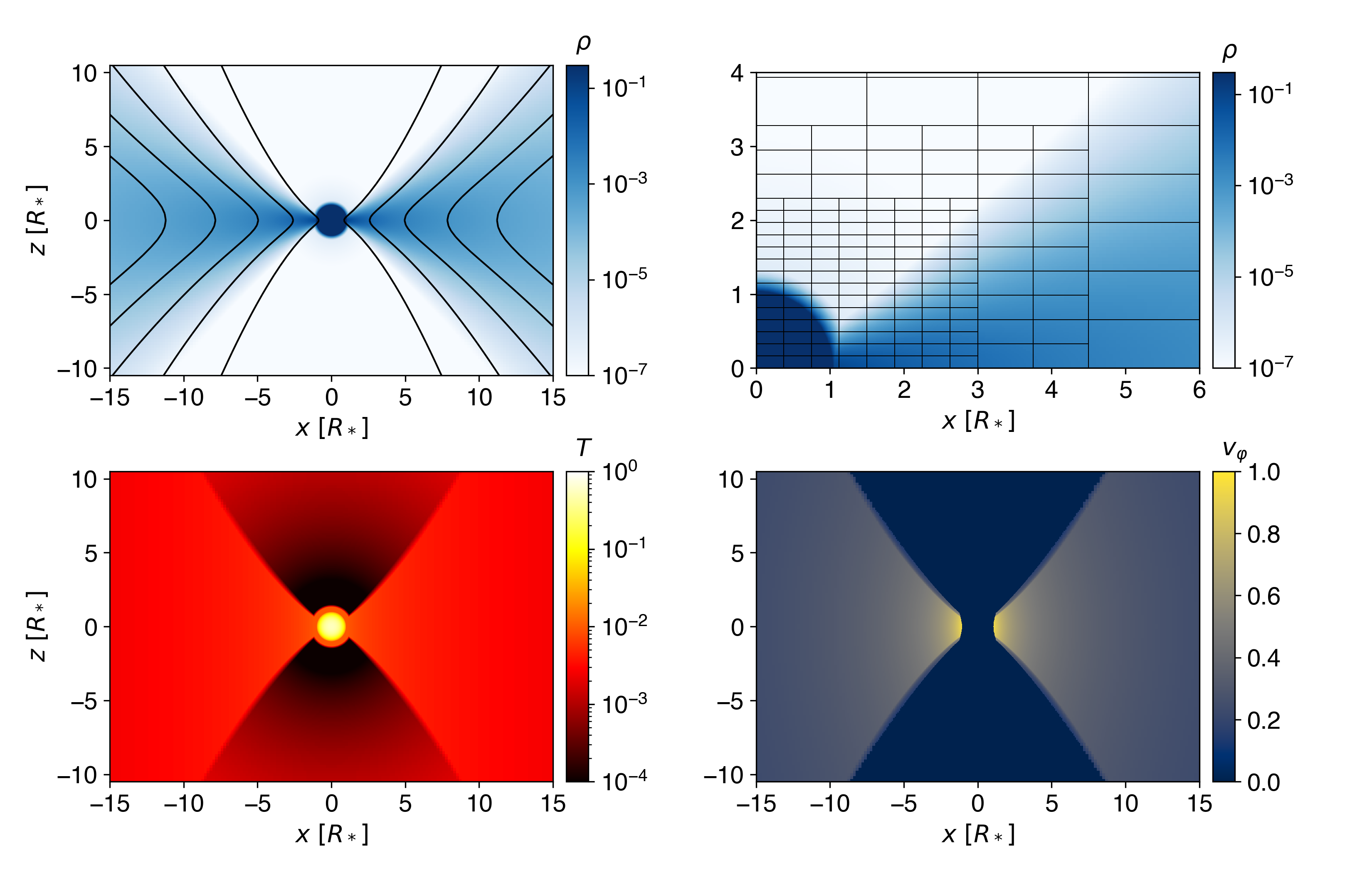}
    \caption{Initial conditions of our model. Top left: the density map with magnetic field lines. The mass density unit is $\rho_0 = 2.0 \times 10^{-7}~{\rm g~cm^{-3}}$. Top right: a zoom-in view of the density map. The black lines indicate the boundaries between Meshblocks, each of which is resolved by $32\times 14$ grids in the $xz$ plane. Bottom left: the temperature map. The temperature unit is $T_0 = 2.1 \times 10^6~{\rm K}$. Bottom right: the azimuthal velocity map. The speed is normalized by the Keplerian velocity at the protostellar radius. For more details about the normalization units, see Section~\ref{subsec:units}.}
    \label{fig:ic_setup}
\end{figure*}

Our initial setup includes a protostar surrounded by a rotating disk. Figure~\ref{fig:ic_setup} displays the density, the temperature and the azimuthal velocity component in the initial condition. We will refer to this figure in the following. Below, the gas pressures for the protostellar and disk components are denoted by $p_\ast$ and $p_{\rm d}$, respectively. To ensure continuity in gas pressure, we adopt the protostellar profile where $p_\ast > p_{\rm d}$, and the disk profile elsewhere. These components are constructed separately as described later. We express physical quantities in spherical and cylindrical coordinates in some figures, but we note that the simulation itself is conducted in Cartesian coordinates.

\subsubsection{Protostar}\label{subsec:protostar}

The protostar is modeled as a non-rotating, hydrostatic polytropic gas sphere with a convectively unstable envelope. The protostellar mass and radius are denoted as $M_\ast$ and $R_\ast$, respectively, and are set to $M_\ast = 0.5M_\odot$ and $R_\ast = 5.5R_\odot$.

This protostellar model is motivated by the results of a 1D protostellar evolution simulation using the code of \citet{Hosokawa2011ApJ}. The simulation begins with an initial mass of $0.05M_\odot$ and an accretion rate of $10^{-4}M_\odot~{\rm yr}^{-1}$. The outer boundary condition controls the entropy injection history. Until the protostar reaches a mass of approximately $0.3M_\odot$, we adopt an accretion shock boundary \citep{Stahler1980}. After this point, a photospheric boundary is used to simulate disk accretion, allowing the protostar to lose heat through radiation \citep{Palla1992ApJ, Hosokawa2010ApJ}. This process results in development of a convective envelope around a central radiative core. By the time the protostar reaches $0.5M_\odot$, its radius is approximately $5.5R_\odot$. Based on this result, we adopt these values for the protostellar mass and radius in our model.
However, we do not directly import the 1D evolution model data into the 3D simulation. Instead, we modify the stellar structure to ensure the feasibility of the simulation (details are provided below).

The density and pressure vary drastically within the protostar, making it challenging to accurately resolve the realistic stellar structure. To focus on the star-disk interaction region, we artificially reduce the gravity near the center of the protostar. The spherical radial component of the gravitational acceleration, $-g(r)$, is given by:
\begin{align}
    g(r)=
    \begin{dcases}
        \frac{GM_\ast}{r^2} & (r\ge r_a), \\
        \frac{GM_\ast}{r_a^3} r & (r < r_a),
    \end{dcases}\label{eq:grav}
\end{align}
where $r_a$ is the radius at which the functional form changes. We neglect the self-gravity of the accreting gas. As the stellar mass is concentrated near the center, the $g(r) \propto r^{-2}$ profile is accurate near and outside the stellar surface. Equation~\eqref{eq:grav} indicates that $g(r)$ exhibits a sharp peak at $r=r_a$. To prevent numerical oscillations, we smooth this peak over a width of $0.03R_\ast$.

The full protostar is placed in the numerical domain. In order to set up a hydrostatic equilibrium, the temperature profile within the protostar is calculated by numerically integrating the following equation from the protostar(s surface toward the center:
\begin{align}
    \frac{dT}{dr}=-\frac{\mu}{(m+1)R_{\rm gas}}g(r), \label{eq:dTdr}
\end{align}
where $\mu$ is the mean molecular weight, $R_{\rm gas}$ is the gas constant and $m$ is the polytropic index ($p \propto \rho^{1+1/m}$). We adopt $m=1.48$ for the convectively unstable envelope and $m=3$ for the inner convectively stable region. The polytropic index transitions at $r=r_{\rm rad} (<R_\ast)$, where $r_{\rm rad}$ is set larger than $r_a$ to ensure that the gravity modification does not affect the convective layer. In this study, we use $r_a=0.55R_\ast$ and $r_{\rm rad}=0.6R_\ast$.

The stellar surface temperature $T_{\rm ps}$ is set to $2\times 10^4~{\rm K}$, higher than the value predicted by the protostellar evolution model ($\sim 4.8\times 10^3~{\rm K}$). This higher value artificially increases the pressure scale height at the stellar surface, $H_{\rm ph}$, allowing it to be marginally resolved by the grid scale.
We note that the stellar surface convection pattern will depend on the scale height.

The protostar is assumed to have a warm, hydrostatic, non-rotating atmosphere outside its surface (corona). The temperature profile for $r > R_\ast$ is defined as:
\begin{align}
    T_\ast(r) = T_{\rm ps} + 0.5(T_{\rm co} - T_{\rm ps}) \left[\tanh{\left(\frac{r - r_{\rm co}}{w_{\rm co}}\right)} + 1\right],
\end{align}
where $T_{\rm co} = 10T_{\rm ps}$, $r_{\rm co} = 2.5R_\ast$, and $w_{\rm co} = 5H_{\rm ph}$. The protostar appears as the central hot sphere in the temperature map of Figure~\ref{fig:ic_setup}. Using the gravitational acceleration and temperature profiles, we numerically integrate the hydrostatic equation to obtain the density profile:
\begin{align}
    \frac{d\ln\rho}{dr} = -\frac{m}{m+1}\frac{\mu g(r)}{R_{\rm gas}T(r)}.
\end{align}
The stellar surface density is chosen such that the surface gas pressure $p_{\rm ps}$ is close to the value from the evolution model. The model predicts $p_{\rm ps} \sim 10^5~{\rm erg~cm^{-3}}$, but we adopt $p_{\rm ps} = 1\times 10^6~{\rm erg~cm^{-3}}$ to reduce the density contrast within the protostar, facilitating numerical simulations.

\subsubsection{Circumstellar disk}
We construct the axisymmetric disk profile based on the analytic solution of \citet{Nelson2013MNRAS}. The temperature is assumed to be constant along the $z$ axis and depends only on the cylindrical radius $R$. The temperature and midplane density $\rho_{\rm d,mid}$ follow power-law distributions:
\begin{align}
    T_{\rm d}(R) &= T_{\rm d,0} \left(\frac{R}{R_\ast}\right)^{n_{\rm T}}, \\
    \rho_{\rm d,mid}(R) &= \rho_{\rm d,0} \left(\frac{R}{R_\ast}\right)^{n_{\rm d}}, \label{eq:rho_mid}
\end{align}
where $\rho_{\rm d,0}$ and $T_{\rm d,0}$ are the density and temperature at $R=R_\ast$. We adopt $n_{\rm d} = -1.875$ and $n_{\rm T} = -0.75$ as the power-law indices, with $\rho_{\rm d,0} \approx 2.0\times 10^{-8}~{\rm g~cm^{-3}}$ and $T_{\rm d,0} \approx 2.1\times 10^4~{\rm K}$.

The density $\rho_{\rm d}(R, z)$ and the azimuthal velocity component in the poloidal plane $v_{\varphi,\rm d}(R, z)$ are given by:
\begin{align}
    &\rho_{\rm d}(R,z) = \rho_{\rm d,mid}(R) \nonumber \\
    &\times \exp{\left[ 
    \frac{GM_\ast}{c_{\rm iso,d}^2} \left(\frac{1}{\sqrt{R^2+z^2}}-\frac{1}{R}\right) \right]}, \\
    &v_{\varphi,\rm d}(R,z) = v_K(R) \nonumber \\
    &\times \left[(n_{\rm d}+n_{\rm T})\left(\frac{c_{\rm iso,d}^2}{v_K(R)}\right)^2 + (1+n_{\rm T})-\frac{n_{\rm T}R}{\sqrt{R^2+z^2}}\right]^{1/2},
\end{align}
where $v_K(R) = \sqrt{GM_\ast/R}$ and $c_{\rm iso,d}=\sqrt{R_{\rm gas}T_{\rm d}/\mu}$ is the isothermal sound speed based on the disk temperature $T_{\rm d}$. 
To prevent divergence of $v_{\varphi,\rm d}$ at $R=0$, we multiply $v_{\varphi,\rm d}(R, z)$ by the factor $(R/R_{\rm in})^4$ for $R<R_{\rm in}$, with $R_{\rm in} = 0.3R_\ast$. See Figure~\ref{fig:ic_setup} for the disk structure.

\subsubsection{Initial magnetic fields}
The initial magnetic field is assumed to have an hourglass shape (see the top left panel of Figure~\ref{fig:ic_setup}). Its radial distribution is set such that the plasma $\beta$ remains constant on the disk midplane \citep{Zanni2007A&A,Takasao2018ApJ}. The azimuthal component of the vector potential, $A_\varphi$, is given by:
\begin{align}
    A_\varphi(R,z) &= 
        \frac{B_{z,\rm d}R_\ast}{1+a_{\rm vp}}\left(\frac{R}{R_\ast}\right)^{a_{\rm vp}}\left[ 1 + \frac{1}{m_B^2}\left(\frac{z}{R}\right)^2\right]^{-5/8} \nonumber \\
        &\equiv A_{\varphi,0}(R,z),
\end{align}
for $r > r_{\rm vp}$, and
\begin{align}
    A_\varphi(R,z) = A_{\varphi,0}(R,z_{\rm sph}(R)),
\end{align}
for $r \le r_{\rm vp}$, where $a_{\rm vp} = 1 + (n_{\rm d}+n_{\rm T})/2$. 
We define $z_{\rm sph}(R)$ to ensure that $A_\varphi$ remains constant along the $z$ axis within $r \le r_{\rm vp}$:
\begin{align}
    z_{\rm sph}(R) \equiv \sqrt{r_{\rm vp}^2 - R^2}.
\end{align}
This modification aligns the magnetic field with the $z$ axis within $r = r_{\rm vp}$ to prevent the generation of an infinitely strong field at the center.

The field strength $B_{z,\rm d}$ is chosen such that the plasma $\beta$ on the disk midplane is $10^3$. We adopt $r_{\rm vp} = 1.1r_{\rm rad}$ and $m_B = 0.5$.

\subsection{Cooling term}\label{sec:cooling}

To model the radiative cooling of the disk and protostellar photospheric gases, we adopt a so-called $\beta$-cooling approach. Since we use an artificially high stellar temperature (see Section~\ref{subsec:protostar}), realistic modeling of the thermal structure is beyond the scope of this study. 

The cooling term $\Lambda$ is calculated to evolve the temperature as:
\begin{align}
    \frac{\partial T}{\partial t} =
        \begin{dcases}
        -\frac{T - T_{\rm ref}}{\tau_{\rm cool}} & (T>T_{\rm ref}),\\
        0 & (T \le T_{\rm ref}),
        \end{dcases}
\end{align}
where $T_{\rm ref}$ represents the reference temperature distribution from the star to the disk, and $\tau_{\rm cool}$ denotes the cooling (relaxation) timescale. Note that only cooling is considered.

The reference temperature $T_{\rm ref}$ is constructed by smoothly connecting the protostellar and disk temperature profiles, $T_\ast(r)$ and $T_{\rm d}(R)$, respectively. We describe the functional forms of $T_{\rm ref}$ and $\tau_{\rm cool}$ in Appendix~\ref{app:cooling}.

\subsection{Magnetic diffusivity model}\label{sec:artificial_diff}

The ideal MHD approximation is valid near the protostar due to the sufficiently high ionization degree \citep{Gammie1996ApJ,Desch2015ApJ}. However, numerical simulations often become unstable when magnetic reconnection occurs due to grid-scale diffusion in low-$\beta$ regions. To improve numerical stability, we implement an artificial magnetic diffusivity that activates only in these sharp current sheets.

In our model, sharp electric current sheets form in various regions, such as the protostellar corona above the convective photosphere and the atmosphere above the MRI-active disk. The artificial diffusivity is designed to minimize its influence on high-$\beta$ regions like the disk midplane.

The magnetic diffusivity is expressed as:
\begin{align}
\eta = c_\eta \Delta \xi v_A {\rm min}\left[\left(\frac{\Delta \xi |\bm{J}|}{|\bm{B}|+\epsilon}\right)^{n_{\eta}}, 1\right],
\end{align}
where $c_\eta$ is a nondimensional parameter, which is set to 0.3 in this study. $\Delta \xi$ represents the local grid size (in our simulations, $\Delta \xi = \Delta x = \Delta y = \Delta z$). $|\bm{B}|$ and $|\bm{J}|$ are the magnetic field strength and current density, respectively. The parameter $n_{\eta}$, controlling the sensitivity to the current density, is set to 2. A small value $\epsilon$ ($10^{-10}$ in numerical units) prevents division by zero.

This formulation ensures that the local Lundquist number, $S_{\rm loc}$ (the ratio of the diffusion time to the Alfv\'en time), based on the local grid size $\Delta \xi$, satisfies:
\begin{align}
    S_{\rm loc} = \frac{v_A \Delta \xi}{\eta} \ge c_\eta^{-1}.
\end{align}
The minimum $S_{\rm loc}$ is reached only within sharp current sheets, ensuring that the diffusivity does not significantly affect other regions.

Numerical tests indicate that $c_\eta = 0.1-1$ effectively suppresses numerical instabilities. For this study, we adopt $c_\eta = 0.3$. We note that this formulation is compatible with mesh refinement. The behavior of the artificial diffusivity is demonstrated in Appendix~\ref{app:artificial_diffusivity}.

\subsection{Boundary conditions}\label{subsec:bc}

The top and bottom ($z$) boundaries are quasi-open, limiting incoming flows. For outgoing gas, zero-gradient boundary conditions are applied to the hydrodynamic quantities and $B_z$, while $B_x = B_y = 0$ in the ghost cells. The horizontal magnetic field is set to zero to ensure that the magnetic pressure gradient force acts outward \citep{Lesur2013A&A}.

For incoming gas, the same boundary conditions are applied if the magnitude of $v_z$ is less than $v_{z,\rm lim} = 0.5v_{\rm esc,bnd}$, where $v_{\rm esc,bnd}(x, y, \pm z_{\rm max}) = \sqrt{2GM_\ast/(x^2 + y^2 + z_{\rm max}^2)}$ is the escape velocity at the height of the box $z_{\rm max}$. If the inflow speed exceeds this threshold, $v_z$ is capped at $v_{z,\rm lim}$, and the density is set to $\rho_{\rm bnd}$ to limit the incoming mass flux. The density $\rho_{\rm bnd}$ is determined for a given critical accretion rate $\dot{M}_{\rm bnd}$ as:
\begin{align}
    \rho_{\rm bnd} = \frac{\dot{M}_{\rm bnd}}{4\pi z_{\rm max}^2 v_{\rm esc,bnd}}.
\end{align}
This setup imposes an upper limit on the mass accretion rate through the top and bottom boundaries.

As our study focuses on cases where the disk accretion rate is significantly higher than typical values for classical T Tauri stars (CTTSs; approximately $10^{-9}$--$10^{-7}~M_\odot~{\rm yr^{-1}}$), we set $\dot{M}_{\rm bnd} = 10^{-7}~M_\odot~{\rm yr^{-1}}$ to ensure that accretion from the boundaries does not affect the disk accretion. We confirmed that the inflow mass flux from the top and bottom boundaries remains below $\dot{M}_{\rm bnd}$ in the simulation, as the inflow regions are limited to the areas where the stellar open fields pass through.

The horizontal ($x, y$) boundaries are challenging to handle because the circularly rotating Keplerian disk interacts with the square boundaries. To prevent the loss of the outer disk on a dynamical timescale, we must allow disk gas to flow into the numerical domain. Standard outflow boundaries are insufficient as they cannot control the angular momentum of the incoming gas. Therefore, we adopt a type of quasi-open boundary condition described below.

Consider the horizontal boundary located at $x=x_{\rm min}$, where $x_{\rm min}$ is the minimum value of the $x$ coordinates of the numerical domain. For this boundary, we apply zero-gradient boundary conditions to the magnetic fields and hydrodynamic quantities, except for $v_x$ (the velocity component normal to the boundary). Let $i$ denote the index of the $x$ coordinate and $il$ the index of the active cell adjacent to the boundary. The velocity in the ghost cells is set as:
\begin{align}
    v_{x,i} = \begin{dcases}
        v_{x,il} & (\text{if}~v_{x,il}<v_{\rm in,lim}), \\
        v_{\rm in,lim} & (\text{otherwise}),
    \end{dcases}
\end{align}
where $v_{\rm in,lim}$ is a threshold velocity to prevent runaway growth of incoming flows. For this boundary, $v_{\rm in,lim}$ is defined as a positive value since our disk rotates anti-clockwise in the $xy$ plane. For gas in near-Keplerian rotation, $v_\varphi$ should approximate the Keplerian value. Based on this, the threshold velocity is defined as:
\begin{align}
    v_{\rm in,lim} = v_{\rm K}(r) |\bm{e}_\varphi \cdot \bm{e}_x|,
\end{align}
where $\bm{e}_\varphi$ and $\bm{e}_x$ are the unit vectors in the azimuthal and $x$ directions, respectively.
Note that $v_{\rm in,lim}$ differs at different points on the boundary at $x=x_{\rm min}$ and is relevant only for incoming flows.

The same boundary condition is applied to the other horizontal boundaries. For the boundary at $x=x_{\rm max}$, the sign of $v_{\rm in,lim}$ is flipped as the direction of incoming flows is opposite to that at $x=x_{\rm min}$.

\subsection{Normalization units and numerical grids}\label{subsec:units}

The length and velocity units are defined as $L_0 = R_\ast$ and $v_0 = v_{\rm K}(R_\ast) = (GM_\ast/R_\ast)^{1/2} (\approx 1.3 \times 10^7~{\rm cm~s^{-1}})$, respectively, where $G$ is the gravitational constant. The mass density unit is $\rho_0 = 2.0 \times 10^{-7}~{\rm g~cm^{-3}}$, and the mean molecular weight $\mu$ is set to unity for simplicity. These units are used to calculate the normalization factors for other quantities. The time unit is $t_0 = L_0 / v_0 = 2.9 \times 10^4~{\rm s}$, and the temperature unit is $T_0 = 2.1 \times 10^6~{\rm K}$. The unit of the magnetic field strength is $B_0 = \sqrt{4\pi \rho_0} v_0 = 2.1 \times 10^4~{\rm G}$. The accretion rate unit is $6.1 \times 10^{-3}~M_\odot~{\rm yr^{-1}}$.

The simulation domain spans $-15 \le x/R_\ast \le 15$, $-15 \le y/R_\ast \le 15$, and $-10.5 \le z/R_\ast \le 10.5$ (namely, $z_{\rm max} = 10.5R_\ast$ in Section~\ref{subsec:bc}). Static mesh refinement with five levels (including the root level) is employed. The root grid (level 0) consists of $160 \times 160 \times 112$ cells. Each Meshblock, which is the domain decomposition unit in Athena$++$, is resolved by $32\times 32 \times 14$ cells.
At the finest level, the protostellar radius is resolved with approximately 90 cells ($\Delta x/R_\ast \approx 0.011$). The finest cell size is comparable to the pressure scale height of the protostellar surface, $H_{\rm ph}$.
The top right panel of Figure~\ref{fig:ic_setup} illustrates the Meshblock distribution near the protostar, with black lines indicating boundaries between Meshblocks.

\subsection{Floors and caps for physical quantities}

As described in Section~\ref{subsec:basic_eq}, we use the Boris correction to limit the Alfv\'en speed and relax the severe CFL condition. The Boris correction employs a semi-relativistic formulation that constrains all velocities to a modified speed of light, $c_{\rm B}$. By setting $c_{\rm B}$ smaller than the real speed of light, we mitigate the CFL condition imposed by very high Alfv\'en speeds. 

We set $c_{\rm B} = 6 v_0 (\sim 8 \times 10^2~{\rm km~s^{-1}})$, which is significantly larger than the Keplerian velocity at $r=R_\ast$ ($\sim 130~{\rm km~s^{-1}}$). Since the Boris correction does not affect dense accreting gas or disk winds, it does not alter the accretion dynamics. However, the Boris-HLLD scheme can become unstable when the gas moves near $c_{\rm B}$ \citep{Matsumoto2019ApJ}. To prevent this numerical instability, we cap the fluid velocity at $3 v_0$. As fluid velocities rarely exceed this cap in the simulation, it does not impact the main results.

We also apply floors to the density, pressure, and temperature to enhance numerical stability. The density and pressure floors are defined as $\rho_{\rm fl,0}(r/R_\ast)^{-2}$ and $p_{\rm fl,0}(r/R_\ast)^{-2}$, where $\rho_{\rm fl,0} = 10^{-6} \rho_0$ and $p_{\rm fl,0} = 10^{-13} p_0$. In certain regions, particularly near the strongly magnetized protostellar surface, the density can drop to very small values, leading to the continuous formation of tenuous hot plasmas. To avoid this instability, we cap the temperature at $10T_0 (\sim 2 \times 10^7~{\rm K})$ and impose an additional density floor to ensure that the original Alfv\'en speed (before applying the Boris correction) does not exceed $30 v_0$.

\section{Results}\label{sec:results}

The simulation was conducted until $t \approx 600t_0$, which is significantly longer than the convection turnover timescale. With the root-mean-square velocity in the convective envelope, $v_{\rm rms}$, of the order of $10^{-2}v_0$, the turnover timescale is estimated as $\Delta r_{\rm cv}/v_{\rm rms} \sim 40 t_0 (r_{\rm cv}/0.4L_0)(v_{\rm rms}/10^{-2}v_0)^{-1}$, where $\Delta r_{\rm cv} = 0.4L_0$ is the thickness of the convective envelope.
The net accretion rate in our model is $\sim 4 \times 10^{-4}$, corresponding to $\sim 2.4 \times 10^{-6}~M_\odot~{\rm yr^{-1}}$ in physical units. The net accretion rate remains approximately constant within a radius of $r \simeq 4R_\ast$.

Explosive phenomena become prominent after $t \approx 480 t_0$. 
We use data prior to this time for analyzing the accretion structure.
The explosive events are described in detail in Section~\ref{subsec:explosive}.

\subsection{General structure: accretion and ejection}\label{sec:overview} 
\begin{figure*}
    \centering
    \includegraphics[width=2.0\columnwidth]{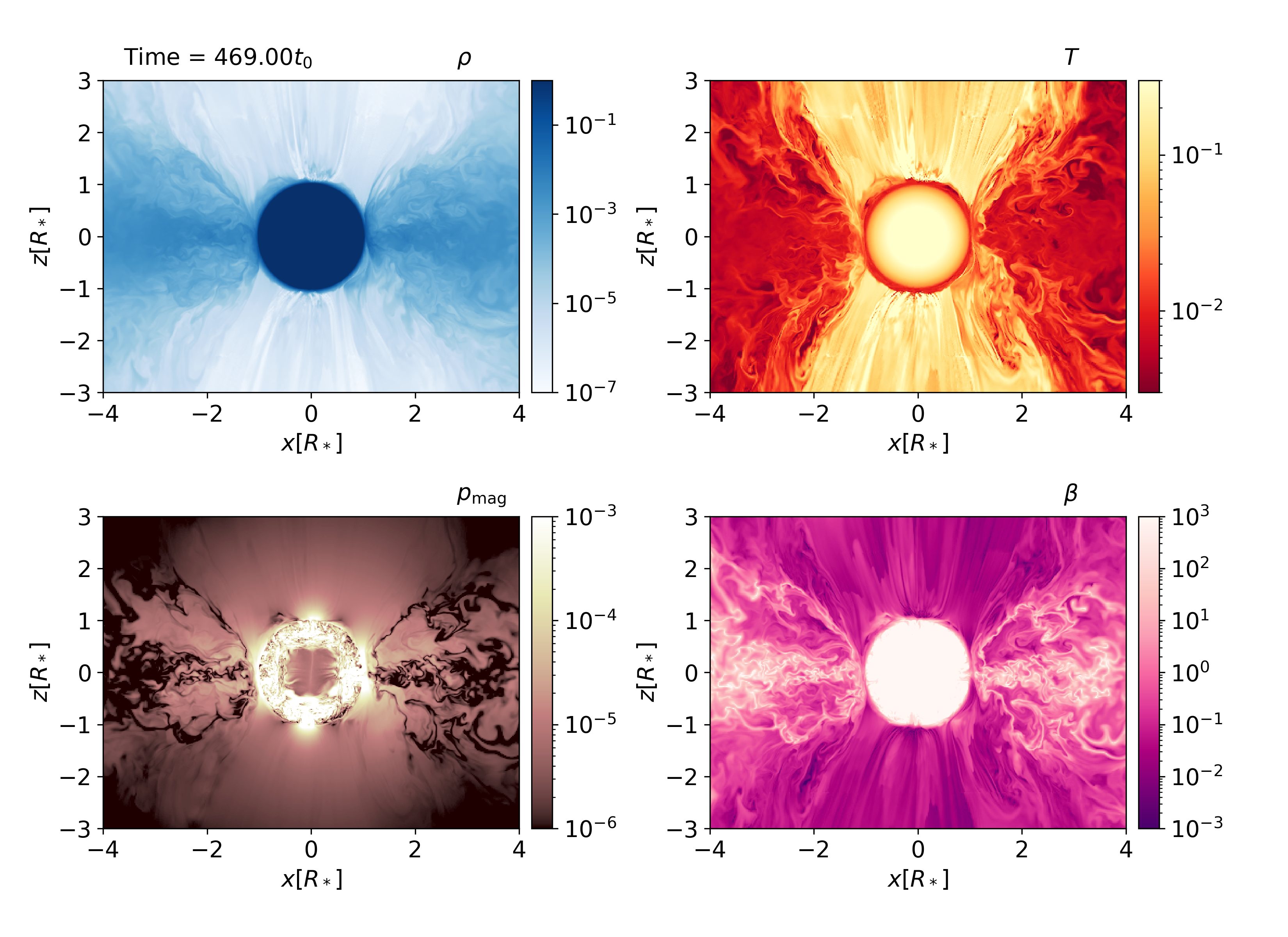}
    \caption{Accretion structure in the $xz$ plane at $y=0$. The panels show (from top left to bottom right) the density, temperature, magnetic pressure, and plasma $\beta$.}
    \label{fig:overview2d}
\end{figure*}

Figure~\ref{fig:overview2d} shows the general gas structure at $t=469 t_0$. The density and temperature maps reveal that the MRI-active accretion disk extends to the protostellar surface, covering the low-latitude regions, while the polar regions host a tenuous, hot corona. The simulation indicates that the protostar acquires disk fields through accretion and accumulates open fields in the polar regions (see the magnetic pressure map, $p_{\rm mag}$). 
The plasma $\beta$ map highlights that low-$\beta$ regions are present not only in the polar regions but also in the disk atmosphere. These low-$\beta$ regions make explosive magnetic energy release possible in the disk atmosphere. This is supported by the temperature map, which shows hot plasmas in the low-$\beta$ disk atmosphere.

The convective motion in the protostellar envelope amplifies the magnetic fields in the stellar interior, as shown in the $p_{\rm mag}$ map. 
In contrast, the central region of the protostar does not amplify the magnetic field because it is non-convective (see Section~\ref{subsec:protostar}).

\begin{figure*}
    \centering
    \includegraphics[width=2.1\columnwidth]{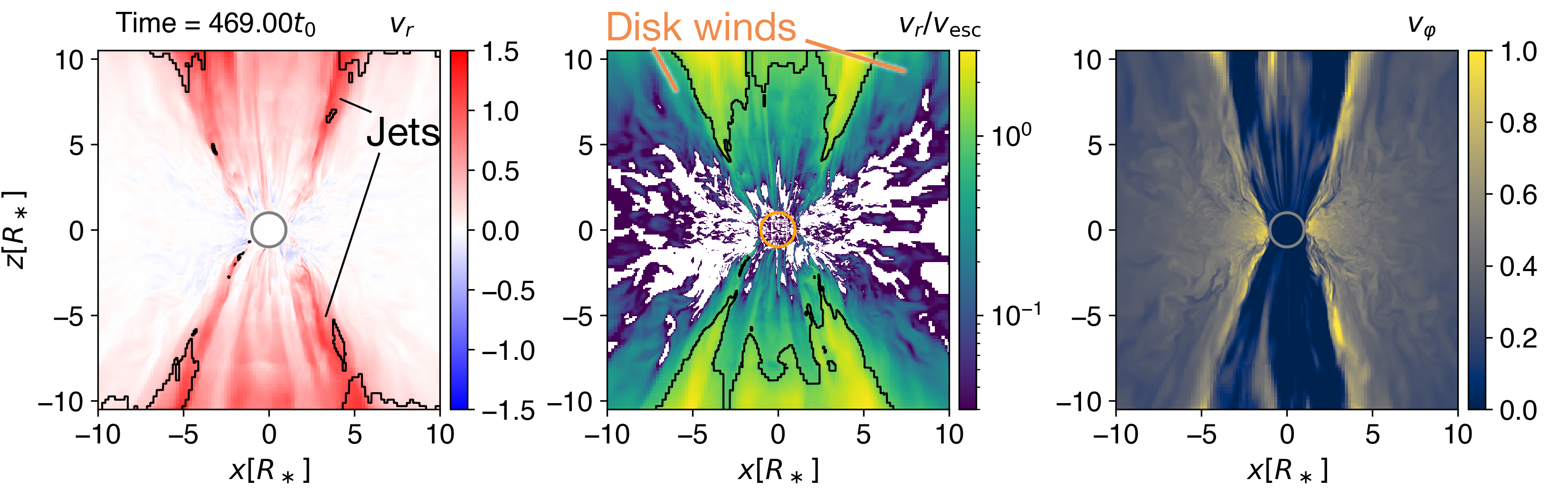}
    \caption{Overall jet and wind structures. The left panel shows the $v_r$ map, with contours representing the approximate fast magnetosonic surfaces ($v_r = \sqrt{v_A^2 + c_s^2}$, where $v_A$ and $c_s$ are the Alfv\'en and adiabatic sound speeds, respectively). The middle panel displays $v_r$ normalized by the local escape velocity, $v_{\rm esc}$, with contours indicating locations where $v_r = v_{\rm esc}$. The right panel presents the azimuthal velocity, $v_\varphi$. In all panels, the central circles represent the stellar surface ($r = R_\ast$).}
    \label{fig:jet}
\end{figure*}

The overall ejection structure is shown in Figure~\ref{fig:jet}. The $v_r$ map reveals a collimated bipolar jet emerging from the star-disk boundary. 
The left panel displays $v_r$, highlighting the collimated bipolar jets.
The middle panel ($v_r$ normalized by the local escape velocity $v_{\rm esc}$) indicates that the jet is escaping the protostellar gravity. 
The right panel ($v_\varphi$ map) shows that the jet is rotating, suggesting that it extracts angular momentum from the accreting gas. As discussed in Section~\ref{subsec:explosive}, the jet is found to be unsteady. Surrounding the jet is a slow wind with a wider opening angle (see the middle panel of Figure~\ref{fig:jet}), which we refer to as the disk wind.

\begin{figure*}
    \centering
    \includegraphics[width=1.7\columnwidth]{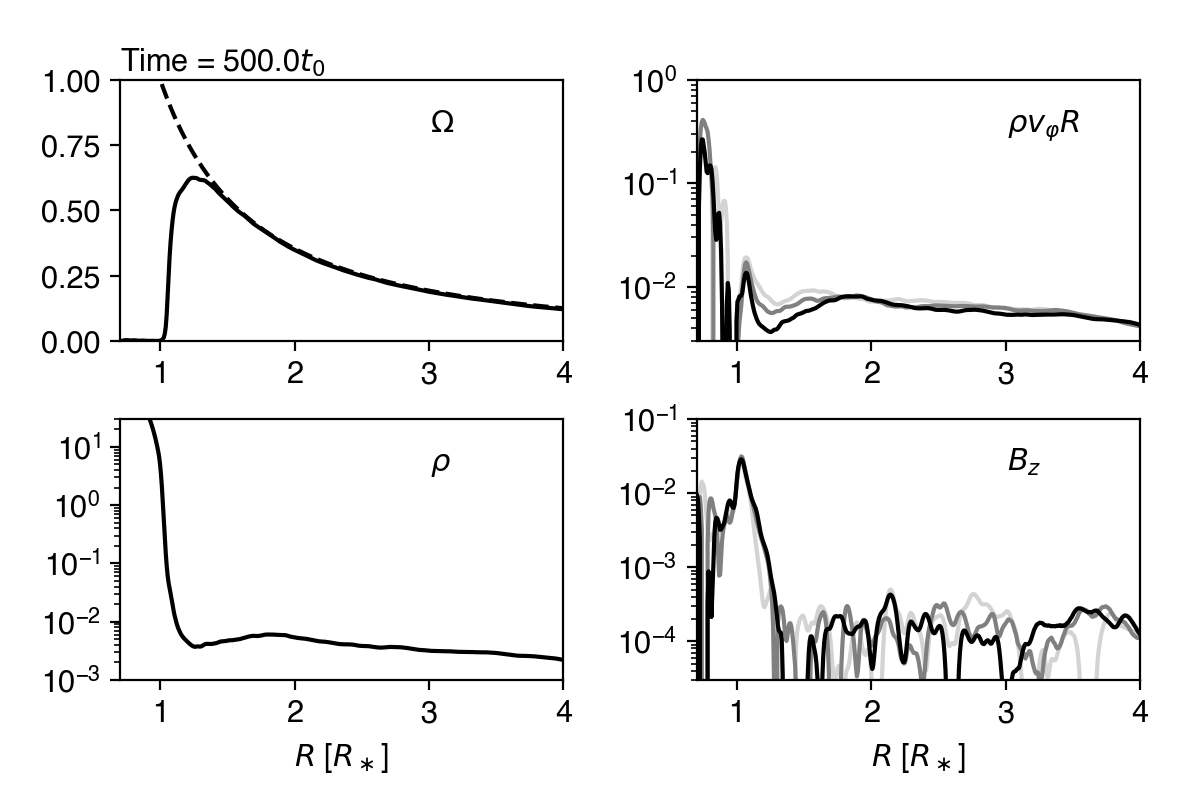}
    \caption{Radial profiles of azimuthally averaged quantities on the equatorial plane at $t=500t_0$. Top left: angular velocity. The dashed line represents the Keplerian profile. Top right: angular momentum density. Bottom left: density. Bottom right: $B_z$. In the angular momentum density and $B_z$ panels, the light gray, gray, and black lines correspond to data at $t=450t_0$, $475t_0$, and $500t_0$, respectively. Data are shown only for $R > 0.7R_\ast$ (the convective layer lies within $r = 0.6R_\ast$).}
    \label{fig:1d_plots}
\end{figure*}

Figure~\ref{fig:1d_plots} presents the radial profiles of azimuthally averaged quantities on the equatorial plane at $t = 500t_0$ (showing data for $R/R_\ast > 0.7$). The top left panel shows the angular velocity $\Omega$, which is used to define the boundary layer. Since the protostar does not rotate initially, $\Omega$ drops sharply to zero near $R = R_\ast$. We define the boundary layer as the region between the protostar and the point where $d\Omega/dr = 0$. Using this definition, the boundary layer is located in the range $1 \lesssim R/R_\ast \lesssim 1.2$ at $t = 500t_0$.

Accreting material carries angular momentum to the protostar, leading to an increase in the angular momentum density, $\rho R v_\varphi$, within the stellar interior (the top right panel of Figure~\ref{fig:1d_plots}). The density profile, shown in the bottom left panel, exhibits a complex structure. From the outer disk toward the protostar ($R/R_\ast \gtrsim 2$), the density approximately follows the initial power-law profile. However, it decreases toward the protostar in the region $1.2 \lesssim R/R_\ast \lesssim 2$ and increases again within the boundary layer. This complex density structure suggests that multiple angular momentum transfer mechanisms are at work, which will be discussed in detail in Section~\ref{subsec:angmom}. 
As described in Section~\ref{subsec:explosive}, the density in the inner disk start to decline rapidly after explosive events driven by magnetic reconnection become prominent. The continuous decrease in the angular momentum density in the boundary layer is partly due to this effect.
The $B_z$ profile in the bottom right panel shows a peak near the protostellar surface, indicating the accumulation of vertical magnetic fields.



Regarding angular momentum transfer near the protostar, we find the formation of spiral shocks. Their role in angular momentum transport will be discussed in Sections~\ref{subsec:angmom} and \ref{subsec:spiral_shocks}.

\subsection{General structure: magnetic fields}\label{sec:overview_mag} 

\begin{figure}
    \centering
    \includegraphics[width=\columnwidth]{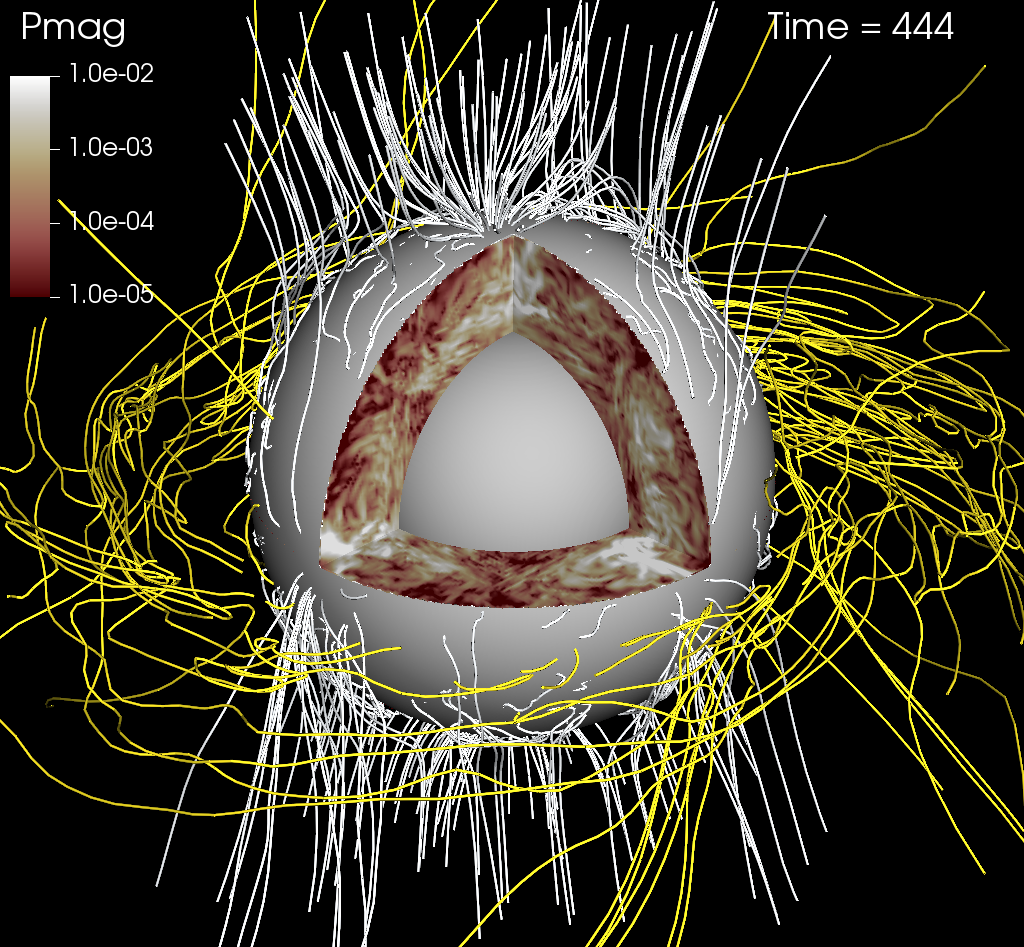}
    \caption{Magnetic structure around the protostar. The outer spherical surface represents the protostellar surface ($r=R_\ast$), while the inner spherical surface corresponds to the bottom of the convective layer ($r=0.6R_\ast$). The colors of these surfaces are for visualization purposes only and carry no physical meaning. Yellow lines depict the disk toroidal magnetic fields, and gray lines represent the field lines that intersect the protostellar surface. The color within the protostar indicates the distribution of magnetic pressure.}
    \label{fig:magnetic_structure3d}
\end{figure}

Figure~\ref{fig:magnetic_structure3d} provides a 3D view of the magnetic structure. The sides of the protostar are wrapped by toroidal fields originating from the rotating disk gas, while the polar regions are dominated by open poloidal fields. By $t \approx 500t_0$, the protostar's unsigned magnetic flux at the stellar surface increases by a factor of two due to the combined effects of the stellar dynamo and the accumulation of the disk fields. Analysis shows that the polar open fields are primarily composed of the disk-origin fields.

At $t\approx 500 t_0$, the protostellar unsigned flux is approximately $2 \times 10^{26}~{\rm Mx}$, and the average stellar field strength is $\sim 110~{\rm G}$. A comparison between the 3D magnetic structure (Figure~\ref{fig:magnetic_structure3d}) and the 2D maps in Figure~\ref{fig:overview2d} indicates that the polar tenuous coronae form in the regions dominated by the open fields. This highlights the critical role of magnetic geometry in shaping the gas structure around the protostar.

\begin{figure}
    \centering
    \includegraphics[width=\columnwidth]{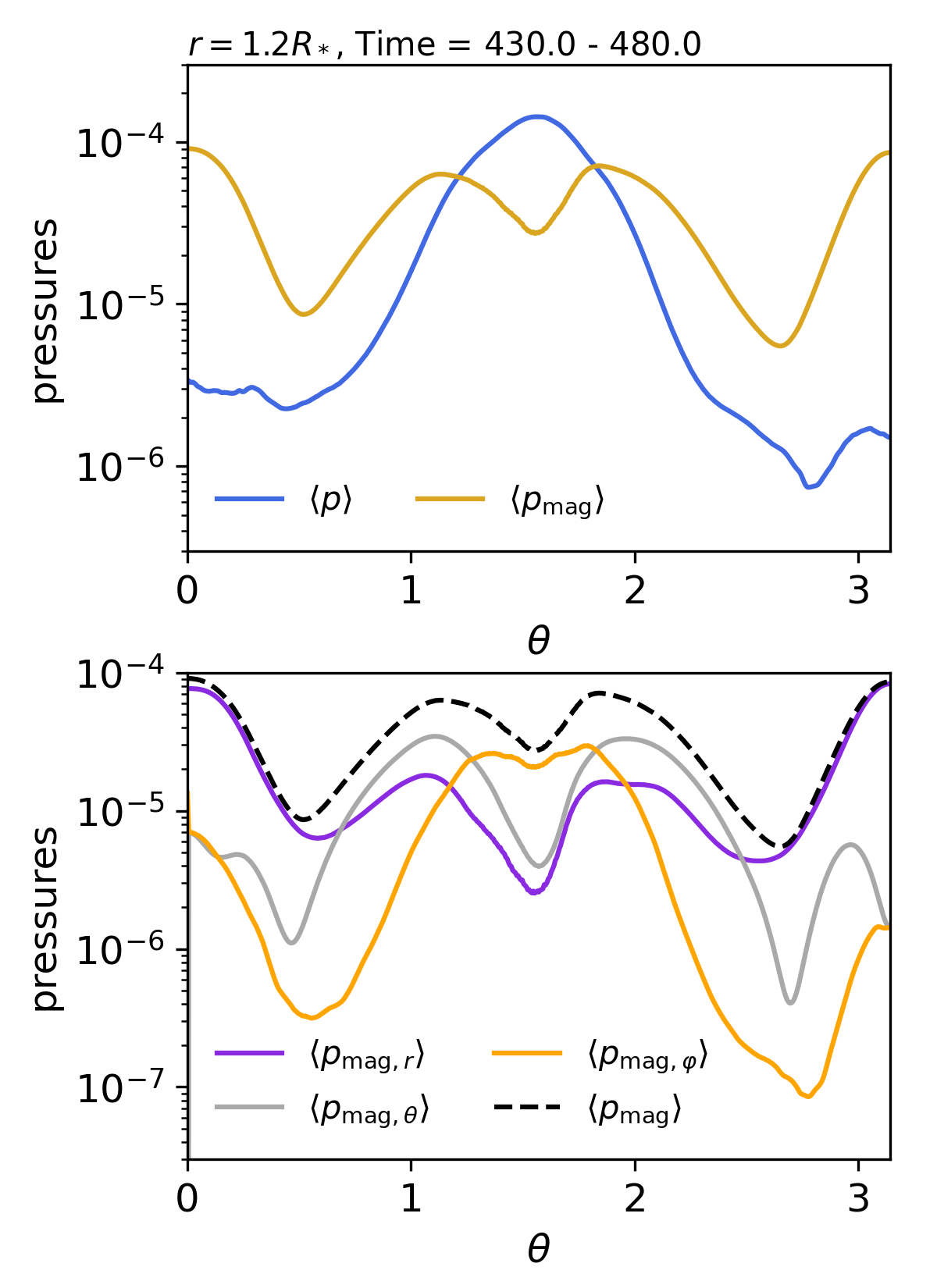}
    \caption{Latitudinal distributions of the gas and magnetic pressures near the protostellar surface ($r=1.2R_\ast$). The data are azimuthally averaged over the period of $430$--$480t_{0}$. Top panel: gas pressure (blue) and total magnetic pressure (gold). Bottom panel: magnetic pressures corresponding to the $r$ (purple), $\theta$ (gray), and $\varphi$ (orange) components of the magnetic field. The dashed line indicates the total magnetic pressure.}
    \label{fig:pressure_balance}
\end{figure}

Figure~\ref{fig:pressure_balance} shows the latitudinal distributions of gas and magnetic pressures near the protostellar surface. The top panel illustrates the gas pressure and total magnetic pressure, revealing that magnetic pressure dominates gas pressure outside the range $1.2 \lesssim \theta \lesssim 2$. The height where gas and magnetic pressures balance roughly corresponds to the disk pressure scale height. The plot indicates that the magnetic field strength around the protostar is governed by pressure balance, with the magnetic pressure constrained by the disk surface gas pressure. This pressure balance is consistent with our previous simulation \citep{Takasao2019ApJ}, where the protostar was not resolved and treated as the inner boundary of the simulation.

The bottom panel of Figure~\ref{fig:pressure_balance} displays the magnetic pressures associated with the $r$, $\theta$, and $\varphi$ components of the magnetic field. The toroidal field dominates in low-latitude regions, while the poloidal field dominates at higher latitudes (see also Figure~\ref{fig:magnetic_structure3d}). The dips in magnetic pressure around $\theta = 0.5$ and $\theta = 2.6$ result from funnel accretion. This process drags disk fields, creating current sheets where the radial and toroidal components of the magnetic fields flip their signs. The characteristics of the funnel accretion are discussed in Sections~\ref{subsec:angmom} and \ref{subsec:slow_funnel}.

\begin{figure*}
    \centering
    \includegraphics[width=2.1\columnwidth]{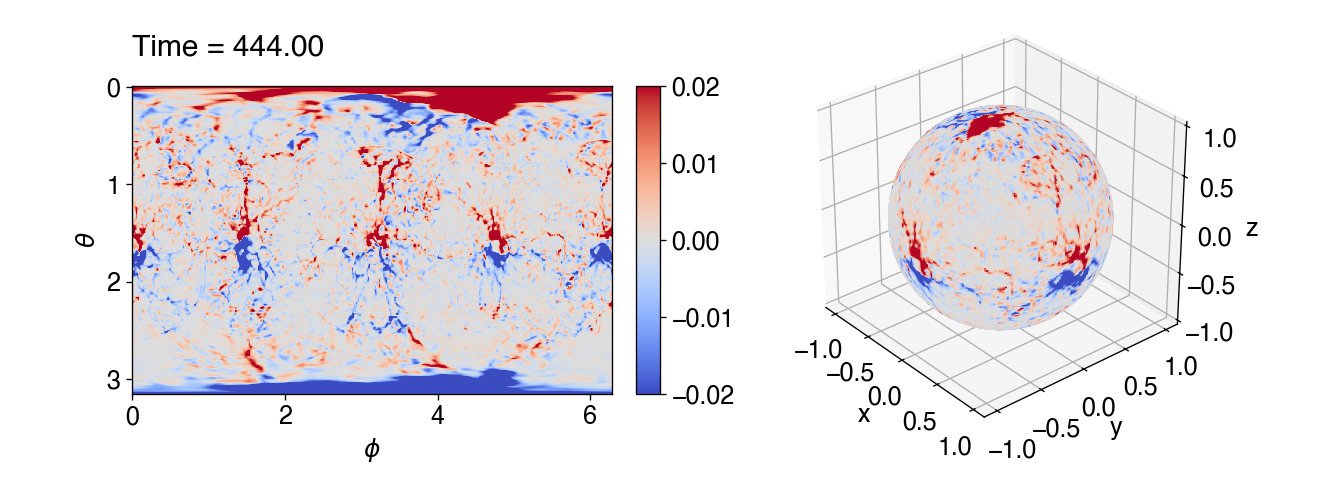}
    \caption{The distributions of the protostellar magnetic fields on the protostellar surface. The radial component $B_r$ at $r=R_\ast$ is displayed. Left: distribution in the rectangular $\theta$-$\varphi$ plane. Right: spherical projection.}
    \label{fig:Brad_sphslc}
\end{figure*}

Magnetic fields at the stellar surface are concentrated in patchy regions. Figure~\ref{fig:Brad_sphslc} shows distributions of the radial magnetic field component, $B_r$, on the protostellar surface. For the data analysis, we utilized the geodesic grid infrastructure developed by \citet{Daszuta2021ApJS} and \citet{White2023ApJ}. The map reveals large magnetic concentrations and network structures, with the network approximately reflecting the convective pattern.
Although the average strength of the stellar magnetic field is $\sim 110~{\rm G}$, magnetic concentrations reach kilo-gauss levels.

The magnetic geometry of the protostar differs significantly from those of CTTSs. Observations of CTTSs often indicate large magnetic arcades or loops with sizes on the order of $R_\ast$ \citep[e.g.,][]{Johnstone2014MNRAS}. In contrast, our protostar lacks well-developed magnetic arcades, despite the continuous generation of magnetic fields inside the protostar and their emergence at the surface.
Figure~\ref{fig:magnetic_structure3d} illustrates how the magnetic field lines within the compact field concentration at the north pole expand dramatically as they extend away from the protostar, but they do not reach the disk. In Section~\ref{subsec:explosive}, we will discuss the role of magnetic reconnection in shaping the protostellar magnetic geometry.

\subsection{Structure of angular momentum flow}\label{subsec:angmom}

\begin{figure*}
    \centering
    \includegraphics[width=2.1\columnwidth]{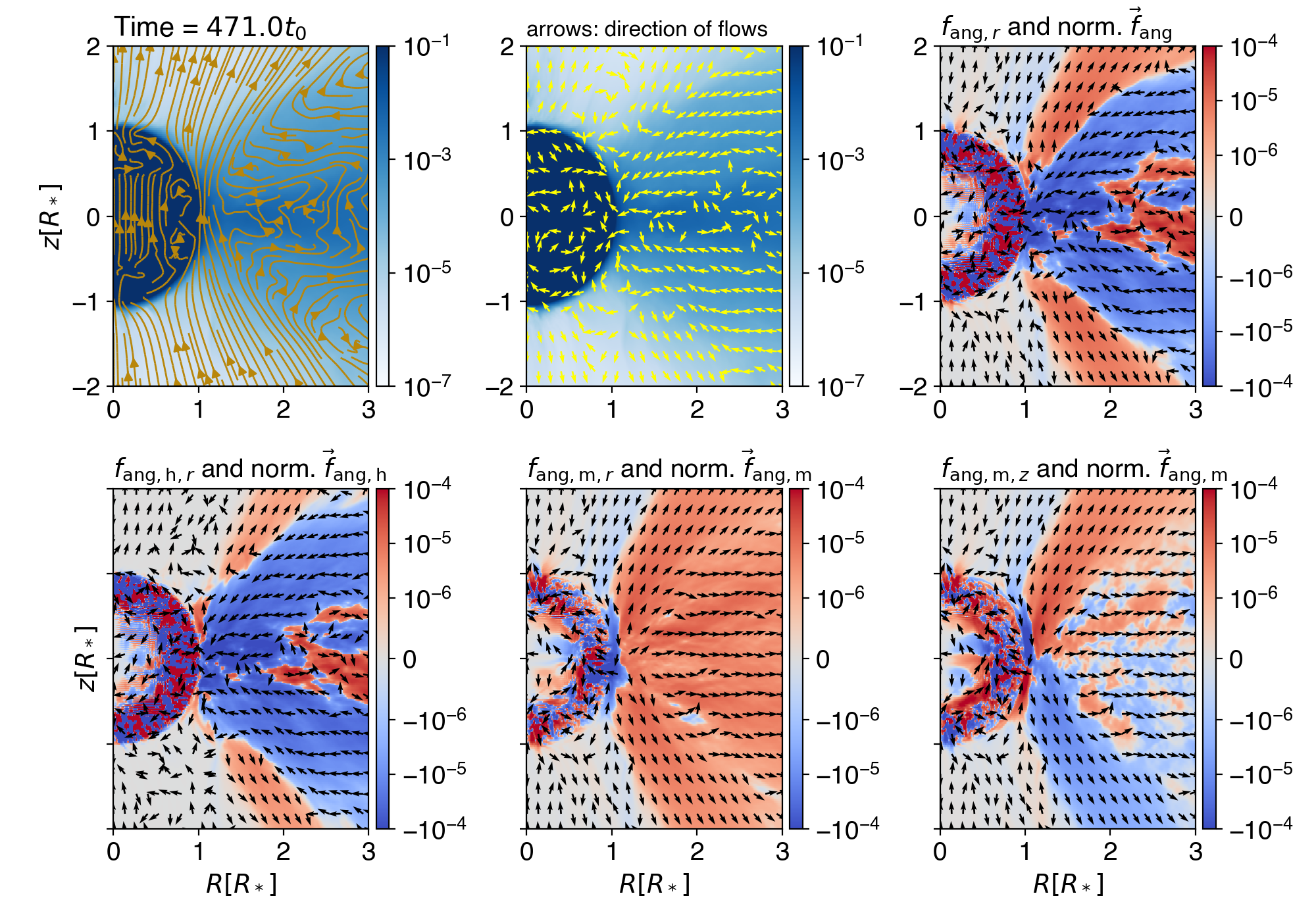}
    \caption{Flow of angular momentum and its relationship to the magnetic and gas flow structures. The data are azimuthally averaged. Top left: density with poloidal field lines overlaid. Top middle: density with arrows representing the direction of gas flows (arrow sizes do not reflect speed). Top right: $r$ component of the total angular momentum flux, $f_{{\rm ang}, r}$, with arrows showing the direction of $\bm{f}_{\rm ang}$. Bottom left: color indicates $f_{{\rm ang,h},r}$, and arrows show the direction of $\bm{f}_{\rm ang,h}$. Bottom middle: color represents $f_{{\rm ang,m},r}$, and arrows denote the direction of $\bm{f}_{\rm ang,m}$. Bottom right: color indicates $f_{{\rm ang,m},z}$ (vertical component), and arrows show the direction of $\bm{f}_{\rm ang,m}$.}
    \label{fig:angflux}
\end{figure*}

Our simulation reveals the presence of MRI turbulence, coronal accretion, disk winds, jets, and spiral shocks, all of which play significant roles in angular momentum transfer. Since angular momentum transfer depends on the magnetic and gas flow structures, we first examine these structures using azimuthally averaged data.

The boundary layer is located within $1 \lesssim R/R_\ast \lesssim 1.2$. To study the transition from the MRI-active disk to the boundary layer, we focus on the region within $r \simeq 4R_\ast$, where the radial profile of the net accretion rate remains approximately constant.

To analyze the angular momentum flow, we define the following angular momentum fluxes:
\begin{align}
    \bm{f}_{\rm ang, h}&=R \rho v_\varphi \bm{v}_p,\\
    \bm{f}_{\rm ang, m}&=-\frac{1}{4\pi}R B_\varphi \bm{B}_p,\\
    \bm{f}_{\rm ang}&=\bm{f}_{\rm ang, h}+\bm{f}_{\rm ang, m},
\end{align}
where $\bm{f}_{\rm ang, h}$ and $\bm{f}_{\rm ang, m}$ represent the angular momentum flux vectors associated with the Reynolds and Maxwell stresses, respectively, and $\bm{f}_{\rm ang}$ is the net angular momentum flux vector. Here, $\bm{v}_p$ and $\bm{B}_p$ denote the poloidal velocity and magnetic field vectors, respectively.

Figure~\ref{fig:angflux} illustrates the structures of the poloidal magnetic fields, gas flow, and angular momentum flux. The directions of these vectors are shown from the top right to the bottom right panels, including the $r$ and $z$ components of the angular momentum flux vectors.

The top left panel of Figure~\ref{fig:angflux} shows that disk fields are dragged toward the protostar outside the disk body, forming a typical structure resulting from coronal accretion. The locations of current sheets in these dragged fields correspond to dips in the magnetic pressure around $\theta \approx 0.5$ and $2.6$ in Figure~\ref{fig:pressure_balance}. Poloidal fields accumulate in the protostellar polar regions and the boundary layer, forming smoothly expanding open field lines.

We can compare the structures of the gas flow and angular momentum flow. The top middle panel of Figure~\ref{fig:angflux} illustrates the gas flow structure, while the top right, bottom left, and bottom middle panels show the radial components of $\bm{f}_{\rm ang}, \bm{f}_{\rm ang,h}$, and $\bm{f}_{\rm ang,m}$, respectively. The gas flow is directed toward the protostar over a wide range of latitudes (approximately within $30^\circ$ from the equatorial plane). Outgoing flows are also observed outside the disk, including two distinct types: the disk wind and the jet from the boundary layer, as described later.

Another notable feature is the ``decretion" region near the equatorial plane, observed in the range $R \gtrsim 1.5\text{-}2R_\ast$. The inner edge of the decretion region varies over time, likely due to the influence of the time-varying stellar fields on the star-disk boundary, as discussed in Section~\ref{subsec:spiral_shocks}. The decreting flows transport angular momentum outward (compare the top middle and bottom left panels).

The bottom left and middle panels highlight the angular momentum fluxes due to Reynolds and Maxwell stresses, respectively. Within $\sim 30^\circ$ of the equatorial plane, the total angular momentum flux closely resembles the Reynolds stress component, emphasizing the importance of hydrodynamic processes such as spiral shocks and decretion. Meanwhile, the Maxwell stress transports angular momentum across a broader solid angle. Around the equatorial plane of the disk, Maxwell stress arises from MRI-driven turbulence, while the coronal mechanism amplifies it in the regions above \citep{Takasao2018ApJ, Zhu2018ApJ}. Maxwell stress in the boundary layer is partly produced by stellar fields.

We investigate the vertical transport of angular momentum, a factor often neglected in previous studies \citep[e.g.,][]{Popham1993ApJ,Armitage2002MNRAS,Belyaev2018MNRAS}. The bottom right panel of Figure~\ref{fig:angflux} shows the vertical component of the magnetic term, $f_{{\rm ang,m},z}$. A strong vertical flow of angular momentum is observed from the boundary layer, highlighting the importance of global 3D MHD simulations. This angular momentum is carried by the rotating jet driven from the boundary layer, which accumulates magnetic flux (Figure~\ref{fig:1d_plots}).

Notably, the angular momentum flux is non-zero within the stellar interior (see the hydrodynamic component in Figure~\ref{fig:angflux}). Stellar convection redistributes the accreted angular momentum throughout the convective layer. As shown in the top right panel of Figure~\ref{fig:1d_plots}, the stellar interior gains angular momentum over time. The figure also illustrates the time evolution of angular momentum density from $t = 450t_0$ to $500t_0$ ($50t_0$ corresponds to approximately 8 Keplerian orbital periods at the stellar radius). The angular momentum density peaks in the boundary layer, indicating accumulation, but this peak does not grow over time. Stellar convection diffuses the angular momentum within the stellar interior, preventing continuous accumulation in the boundary layer. A similar argument applies to the $B_z$ profile.

Angular momentum accumulation in the boundary layer was also reported in the local model of \citet{Belyaev2018MNRAS}. However, their results show a continuous growth of the peak angular momentum density, while our findings show that the peak remains approximately steady over time. 
This difference likely arises from different assumptions about the stellar atmosphere. In the local model of \citet{Belyaev2018MNRAS}, the stellar surface is modeled as a hydrostatic, convectively stable layer, causing the accreted angular momentum to pile up and form a belt on the surface. 
In our model, stellar convection redistributes angular momentum within the convective layer, preventing continuous accumulation in the boundary layer. Additionally, vertical transport of angular momentum, which is neglected in the local model, further suppresses the growth of angular momentum density. As we discuss in Section~\ref{subsec:angmom2}, vertical transport plays a significant role near the protostellar surface.

\subsection{Anatomy of angular momentum transfer mechanism}\label{subsec:angmom2}

We have demonstrated that various angular momentum transfer processes are at work. To quantify their relative importance in driving accretion, we use equations that relate the accretion rate to different angular momentum transfer mechanisms \citep[see also][]{Iwasaki2024PASJ}.
Compared to analyses that evaluate the radial transfer of mass and angular momentum through the surfaces of cylindrical annuli, this approach has the advantage of explicitly capturing the contributions of both radial and vertical angular momentum transport.

In the following analysis, the azimuthal average of a physical quantity $Q$ is denoted by $\langle Q \rangle$. Depending on the physical dimension of the quantity, we apply either a simple volume-weighted average or a density-weighted average. Specifically, for quantities with dimensions of mass density, momentum density, energy density, and field strength, we use the volume-weighted average. For velocities, we adopt the density-weighted average.

We define the following stresses:
\begin{align}
    \langle W_{R \varphi} \rangle &= \langle \rho v_R \delta v_\varphi \rangle - \frac{1}{4\pi} \langle B_R B_\varphi \rangle, \\
    \langle W_{\varphi z} \rangle &= \langle \rho v_z \delta v_\varphi \rangle - \frac{1}{4\pi} \langle B_z B_\varphi \rangle,
\end{align}
where $\delta v_\varphi = v_\varphi - \langle v_\varphi \rangle$. 

By combining the equation of continuity and the azimuthal component of the equation of motion, and assuming a steady state, we obtain:
\begin{align}
    \langle \rho v_R \rangle = \frac{2}{\Omega^\prime (R, z)} \left[ J_{R \varphi} + J_{\varphi z} - \langle \rho v_z \rangle \frac{\partial}{\partial z} \langle v_\varphi \rangle \right], \label{eq:rhovR}
\end{align}
where
\begin{align}
    \Omega^\prime (R, z) &= \frac{2}{R} \left( \langle v_\varphi \rangle + R \frac{\partial}{\partial R} \langle v_\varphi \rangle \right), \\
    J_{R \varphi} &= -\frac{1}{R^2} \frac{\partial}{\partial R} \left( R^2 \langle W_{R \varphi} \rangle \right), \\
    J_{\varphi z} &= -\frac{\partial}{\partial z} \langle W_{\varphi z} \rangle.
\end{align}

The accretion rate within the height range $-z_b \le z \le z_b$ is expressed as:
\begin{align}
    \dot{M} = 2\pi R \int_{-z_b}^{z_b} \langle \rho v_R \rangle dz.
\end{align}
Using Equation~\ref{eq:rhovR}, we can estimate the accretion rates associated with different angular momentum transfer mechanisms.
In this study, we set $z_b = 2.5R_\ast$ to ensure that both the coronal accretion regions and the disk wind bases are included in the integration.

By integrating the individual terms in Equation~\ref{eq:rhovR}, we obtain the following expression for the predicted accretion rate:
\begin{align}
    \dot{M}_{\rm pred} = \dot{M}_{R \varphi} + \dot{M}_{\varphi z} + \dot{M}_{\rho v_z}, \label{eq:mdot_pred}
\end{align}
where
\begin{align}
    \dot{M}_{R \varphi} &= 4\pi R \int_{-z_b}^{z_b} \frac{1}{\Omega^\prime} J_{R \varphi} \, dz, \\
    \dot{M}_{\varphi z} &= 4\pi R \int_{-z_b}^{z_b} \frac{1}{\Omega^\prime} J_{\varphi z} \, dz, \\
    \dot{M}_{\rho v_z} &= -4\pi R \int_{-z_b}^{z_b} \frac{1}{\Omega^\prime} \langle \rho v_z \rangle \frac{\partial}{\partial z} \langle v_\varphi \rangle \, dz.
\end{align}
If $\Omega^\prime = \Omega_{\rm K}(R)$, these equations reduce to those in \citet{Iwasaki2024PASJ}.

In this analysis, we also decompose the expressions into hydrodynamic and magnetic terms. For instance, the hydrodynamic and magnetic components of $\dot{M}_{R \varphi}$ are denoted as $\dot{M}_{R \varphi, {\rm h}}$ and $\dot{M}_{R \varphi, {\rm m}}$, respectively. 
$\dot{M}_{R \varphi, {\rm h}}$ represents the contribution from fluctuations in the Reynolds stress, including spiral shocks. $\dot{M}_{R \varphi, {\rm m}}$ is associated with MRI turbulence and the coronal mechanism. $\dot{M}_{\varphi z, {\rm m}}$ corresponds to accretion driven by magnetic braking. $\dot{M}_{\rho v_z}$ represents accretion driven by mass ejection.

There is a caveat regarding the interpretation of the vertical transport terms $\dot{M}_{\rho v_z}$ and $\dot{M}_{\varphi z, {\rm m}}$: their values at a specific radius should include contributions from more inner regions, as the disk magnetic fields are inclined outward. However, our calculation simply integrates vertically to estimate the accretion rates.

\begin{figure}
    \centering
    \includegraphics[width=\columnwidth]{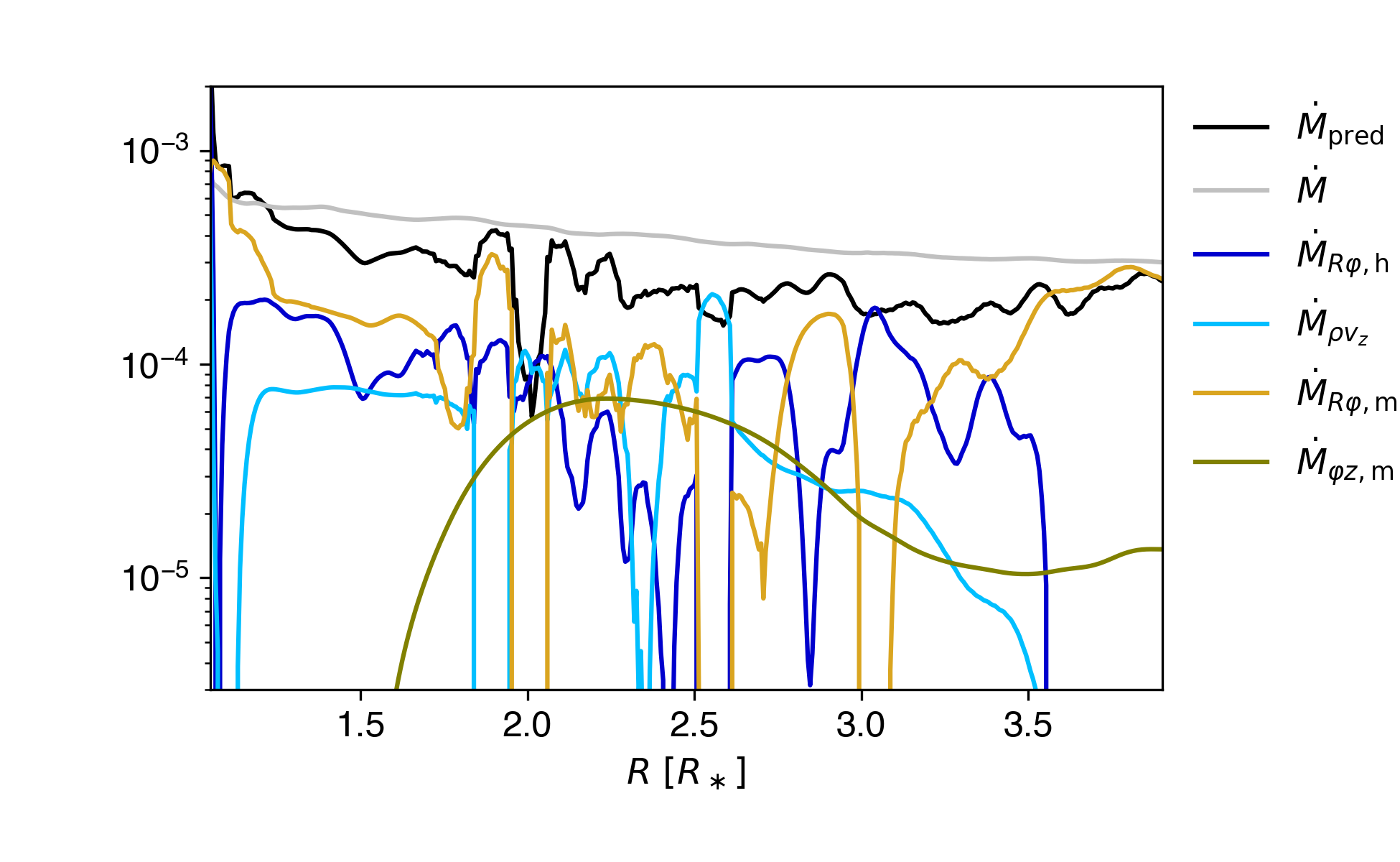}
    \caption{Radial profile of the accretion rates arising from different angular momentum transfer mechanisms. The gray line represents the measured accretion rate $\dot{M}$, while the black line shows the predicted accretion rate $\dot{M}_{\rm pred}$ (Equation~\ref{eq:mdot_pred}). The blue and sky blue lines correspond to $\dot{M}_{R \varphi, {\rm h}}$ and $\dot{M}_{\rho v_z}$, respectively. The orange and olive lines denote the magnetic terms, $\dot{M}_{R\varphi,{\rm m}}$ and $\dot{M}_{\varphi z,{\rm m}}$, respectively. The data are averaged over the period $t=390t_0$--$507t_0$ and smoothed radially. The data in the range $1.05 \le R/R_\ast \le 3.9$ is shown.}
    \label{fig:mdot_from_angmom}
\end{figure}

Figure~\ref{fig:mdot_from_angmom} shows the radial profiles of the accretion rates derived from the different angular momentum transfer mechanisms. To reduce noise in the original data, we present radially smoothed profiles. 
Noticeable fluctuations remain in the plot. We suspect that these fluctuations may arise from the time resolution of the analyzed datasets and the non-stationarity of the system. The time resolution of the analyzed 3D datasets is $3t_0$, which corresponds to approximately half of the Keplerian orbital period at the protostellar radius. The time period for the data averaging, approximately $19t_0$, corresponds to about two orbital timescales at $4R_\ast$. This duration may be insufficient to obtain fully time-averaged data within the radius of $4R_\ast$.

The measured net accretion rate $\dot{M}$ (gray line) and the predicted net accretion rate $\dot{M}_{\rm pred}$ (black line) are similar and nearly constant across radii, confirming the validity of our analysis. The difference between them may be attributed to the non-stationarity, as mentioned above.
For $R \gtrsim 3.3R_\ast$, $\dot{M}_{R \varphi, {\rm m}}$ is the dominant term, indicating that the MRI turbulence and the coronal mechanism primarily drive accretion in this region. In the range $2R_\ast \lesssim R \lesssim 3.3R_\ast$, all four terms contribute approximately equally to the accretion rate.

In the range $1.2 R_\ast \lesssim R \lesssim 2 R_\ast$, $\dot{M}_{R \varphi, {\rm h}}$, $\dot{M}_{R \varphi, {\rm m}}$, and $\dot{M}_{\rho v_z}$ contribute approximately equally to the net accretion rate, although the mass ejection term, $\dot{M}_{\rho v_z}$, is somewhat smaller than the other two. The large value of $\dot{M}_{R \varphi, {\rm h}}$ highlights the significance of spiral shocks in this region.

In the boundary layer ($R\lesssim 1.2 R_\ast$), the magnetic term $\dot{M}_{R \varphi, {\rm m}}$ dominates, highlighting the critical role of sheared magnetic fields. These sheared fields originate from both coronal accretion and the interaction between stellar magnetic fields and the rotating disk gas. A detailed investigation into the relative contributions of these two effects is deferred to future studies. We also note that the spatial resolution of our simulation may be insufficient to reveal the angular momentum transfer mechanisms in the narrow boundary layer.

The simulation highlights the necessity of global 3D MHD models for a comprehensive understanding of angular momentum transfer mechanisms.
Contrary to the expectations of the previous studies, the magnetic term $\dot{M}_{R \varphi, {\rm m}}$ plays an important role even near the protostar. We emphasize the significance of the coronal accretion (including the disk surface and funnel accretion), which drags magnetic fields inward, producing $B_R$ that extends down to the protostellar surface. This behavior is evident in the top left panel of Figure~\ref{fig:angflux}. The differential rotation then generates $B_\varphi$ from $B_R$. Consequently, the $R \varphi$ component of the Maxwell stress works as a significant contributor to angular momentum transfer, even close to the protostellar surface. 
In addition, the boundary layer drives jets, resulting in a substantial contribution from the mass ejection term, $\dot{M}_{\rho v_z}$.

\subsection{Spiral shocks}\label{subsec:spiral_shocks}

\begin{figure*}
    \centering
    \includegraphics[width=2.1\columnwidth]{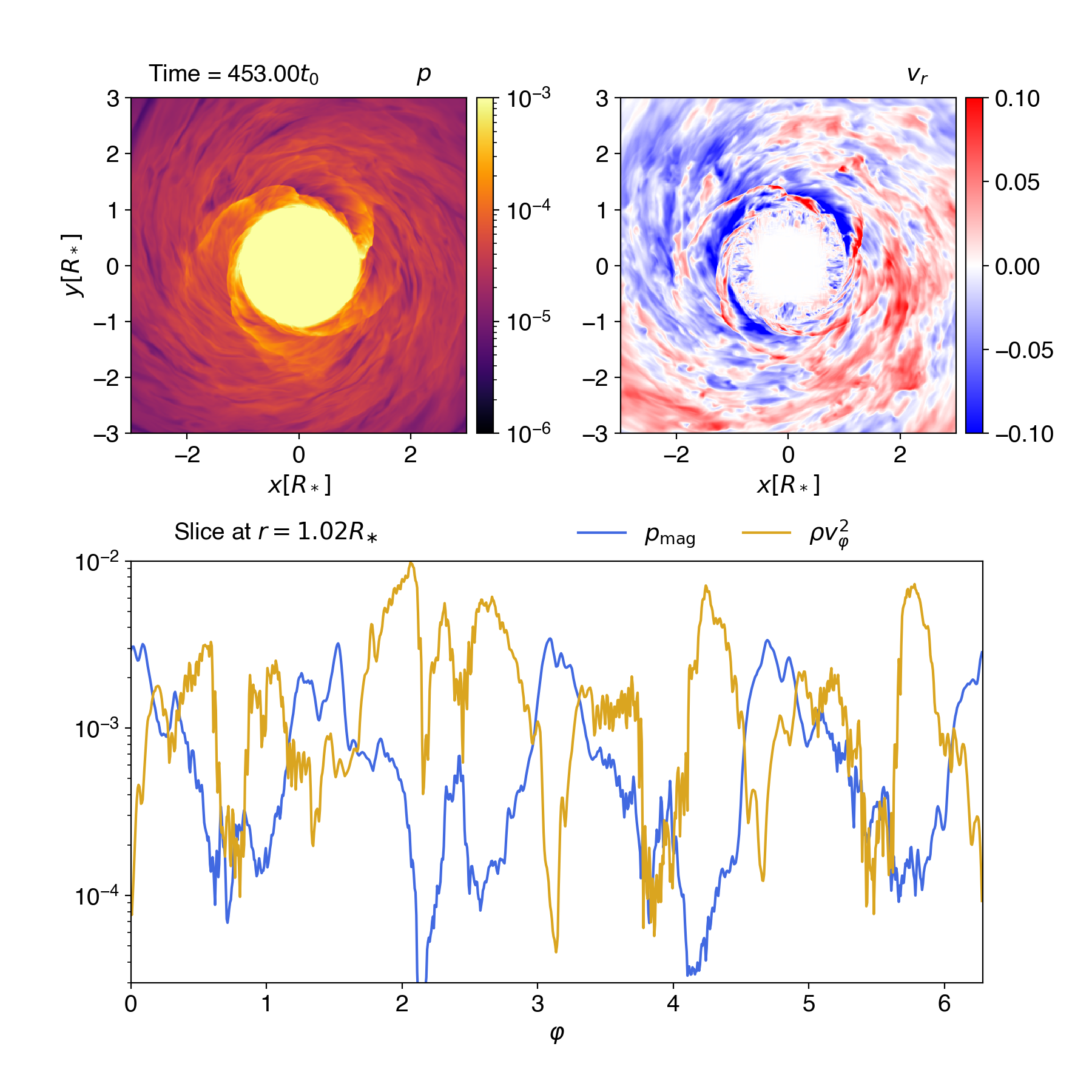}
    \caption{Top left and right: the gas pressure and the radial component of the velocity $v_r$ on the equatorial plane. Bottom: the magnetic pressure ($p_{\rm mag}$) and the ram pressure based on the rotational motion ($\rho v_\varphi^2$) near the protostellar surface. The data are measured along a circle with a radius of $r=1.02R_\ast$ on the equatorial plane at time $453t_0$.}
    \label{fig:spiral_shocks2d}
\end{figure*}

As shown in Section~\ref{subsec:angmom2}, the spiral shocks emanating from the protostellar surface play a crucial role in driving accretion. We investigate their origin, relationship with the protostellar magnetic fields, and contribution to the hydrodynamic term of the accretion rate, $\dot{M}_{R \varphi, \rm h}$.

We begin by examining the structure of the spiral shocks. Figure~\ref{fig:spiral_shocks2d} illustrates the shock structures on the equatorial plane. The top left panel shows the pressure distribution, where spiraling discontinuities correspond to the spiral shocks. Since the disk gas rotates counterclockwise in this plane, the shocks propagate in the opposite direction of the disk rotation. 
The spiral shocks extend to a radius of $\sim 2$--$3R_\ast$ and dynamically change their structures, likely in response to the varying protostellar fields.

The $v_r$ map in the top right panel reveals that the gas just behind the shocks initially moves radially outward before accreting. This outward motion occurs because the flow becomes more parallel to the shocks immediately after their passage. Further away from the shocks, the gas moves inward as its rotational motion is decelerated by the shocks. Given their significant influence on the velocity structure, the spiral shocks are essential drivers of accretion within a radius of a few $R_\ast$.

We find that the spiral shocks in our model are excited near strong magnetic concentrations on the stellar surface. These magnetic concentrations are located around the equatorial plane, with four distinct concentrations observed (see the left panel of Figure~\ref{fig:Brad_sphslc}). This number is consistent with the prominence of the $m_{\varphi}=4$ mode in our simulation (see the top left panel of Figure~\ref{fig:spiral_shocks2d}).This subsection investigates the physical mechanism by which magnetic fields generate spiral shocks. In Section~\ref{sec:model_limitations}, we will discuss the numerical influences on the development of the $m_{\varphi}=4$ mode.

The bottom panel of Figure~\ref{fig:spiral_shocks2d} illustrates the relationship between magnetic pressure and ram pressure due to gas rotation near the protostellar surface ($r=1.02R_\ast$). The plot shows that gas rotation is decelerated as it collides with strong magnetic fields. This collision compresses the gas and excites (fast-mode) magnetosonic waves. The ram pressure near the protostellar surface drops by approximately $100$\%, demonstrating that the protostellar magnetic fields are effective wave exciters. 

The reduction in ram pressure across the shocks is $\mathcal{O}(10^{-3})$ to $\mathcal{O}(10^{-2})$ (Figure~\ref{fig:spiral_shocks2d}), and the post-shock pressure within a radius of $\sim 1.5R_\ast$ is comparable to these values (see the pressure map). MHD waves identified in \citet{Steinacker2002ApJ} may be excited through a similar mechanism, but the exact process remains unclear due to the lack of detailed information in their paper regarding the azimuthal velocity and magnetic field distributions in the boundary layer.

\begin{figure}
    \centering
    \includegraphics[width=\columnwidth]{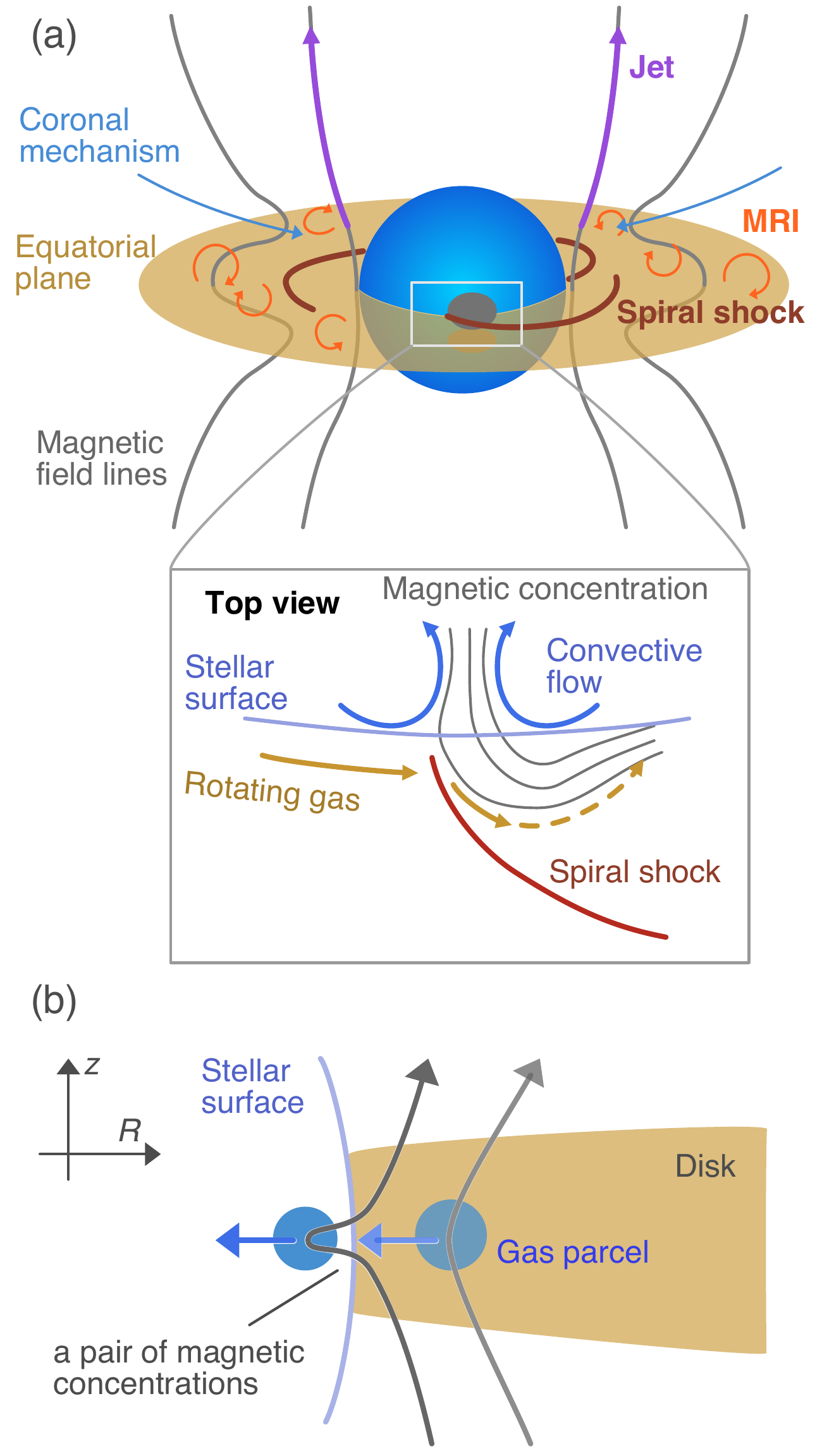}
    \caption{(a) A schematic diagram illustrating the excitation of spiral shocks. (b) An illustration depicting the formation of a pair of magnetic concentrations, showing the evolution of a poloidal magnetic field threading a gas parcel in accretion flows.}
    \label{fig:summary_spiral_shock}
\end{figure}

Figure~\ref{fig:summary_spiral_shock}~(a) presents a schematic diagram illustrating the excitation of the spiral shocks. The zoomed-in panel focuses on the equatorial plane near the protostellar surface, where the protostar primarily receives mass. 
The field strength on the protostellar surface is comparable to the gas pressure, causing the surface to be significantly modulated by the magnetic fields. This modulation creates ``magnetic bumps," making the stellar surface bumpy. The rotating gas collides with these magnetic bumps, exciting large-amplitude magnetosonic waves, which quickly evolve into shocks and form the spiral structures. These spiral shocks extend outward, decelerate the rotating gas within a few stellar radii, and facilitate accretion.

The spiral shocks in our model are magnetically excited, differing from the findings of the previous studies. Some earlier works suggest that spiral shocks in the boundary layer arise from shear-acoustic instabilities \citep[e.g.,][]{Belyaev2013ApJ}. We consider that the magnetically excited spiral shocks will be stronger than their hydrodynamic counterparts, as the stellar magnetic concentrations create more pronounced bumps on the stellar surface in terms of the isosurfaces of total pressure.

The magnetic concentrations near the equator appear as pairs of opposite magnetic polarities (Figure~\ref{fig:Brad_sphslc}). We find the following mechanism (see also the panel~(b) of Figure~\ref{fig:summary_spiral_shock}): as gas accretes onto the protostar through the boundary layer, it drags poloidal fields that have accumulated in the boundary layer (see the $B_z$ plot in Figure~\ref{fig:1d_plots}). This process causes a portion of the poloidal fields to submerge below the stellar surface, forming pairs of opposite polarities.
For example, comparing the $v_r$ map in Figure~\ref{fig:spiral_shocks2d} with the $B_r$ spherical map in Figure~\ref{fig:Brad_sphslc} reveals that the magnetic concentration around $(x, y, z) = (0, -R_\ast, 0)$ forms in the downflow region.
For further discussions on the magnetic field submergence, we refer readers to studies on solar magnetism \citep[e.g.,][]{Cheung2010ApJ,Takasao2015ApJ}.

The hydrodynamic term of the accretion rate, $\dot{M}_{R\varphi, {\rm h}}$, includes the contribution of the spiral-shock-driven accretion but may also contain other effects. To determine whether the spiral shocks dominate this term, we quantify the spiral-shock-driven accretion rate using the theory of \citet{Rafikov2016ApJ}. 

Using the Rankine-Hugoniot relations, \citet{Rafikov2016ApJ} derived the accretion rate as:
\begin{align}
    \dot{M}_{\rm sh} = {\rm sign}[\Omega - \Omega_{\rm p}]\Omega \Sigma H_p^2 m_\varphi \left( \frac{d\ln l}{d\ln R}\right)^{-1} \psi_Q(\Pi),
\end{align}
where $\Omega_{\rm p}$ is the angular velocity of the shock pattern, $H_p = c_{\rm iso} / \Omega$ is the disk's pressure scale height (with $c_{\rm iso}$ being the isothermal sound speed), and $m_\varphi$ is the number of the spiral shocks. The specific angular momentum of the disk fluid is $l = \Omega R^2$. The parameter $\Pi$ represents the ratio of pressures after and prior to passing the shock. \citet{Rafikov2016ApJ} define $\psi_Q(\Pi)$ as a parameter which quantifies the irreversible shock heating per unit mass. The functional form of $\psi_Q(\Pi)$ depends on whether the gas is isothermal or not. Using the effective specific heat ratio $\gamma$, $\psi_Q(\Pi)$ is given by
\begin{align}
    \psi_Q(\Pi) = \frac{1}{\gamma - 1}\left[\Pi \left[\frac{\gamma + 1 + (\gamma - 1)\Pi}{\gamma - 1 + (\gamma + 1)\Pi}\right]^\gamma - 1\right],
\end{align}
for the non-isothermal gas, and 
\begin{align}
    \psi_Q(\Pi) = \frac{\epsilon(2+\epsilon)-2(1+\epsilon)\ln{(1+\epsilon)}}{2(1+\epsilon)},
\end{align}
for the isothermal gas, where $\epsilon\equiv \Pi - 1$.
Considering the effects of disk cooling, we expect the realistic value to lie between these two limits.

To calculate $\dot{M}_{\rm sh}$, we measure physical quantities at $R = 1.5R_\ast$: the density and angular velocity are $\rho \sim 5\times 10^{-3}\rho_0$ and $\Omega \sim 0.6$ (see Figure~\ref{fig:1d_plots}). The number of the spiral shocks varies but is approximately $m_\varphi = 4$ most of the time (Figure~\ref{fig:spiral_shocks2d}), which we use as a fiducial value. The pressure ratio is $\Pi \approx 3$--$4$ (Figure~\ref{fig:spiral_shocks2d}). For $\Pi = 4$, $\psi_Q \sim 2\times 10^{-1}$ and $5\times 10^{-1}$ for the adiabatic and isothermal gas, respectively. We take these values to argue the possible range for a realistic accretion rate. The specific angular momentum $l$ is approximated by the Keplerian value at $R = 1.5R_\ast$, giving $(d\ln l / d\ln R)^{-1} \approx 2$. Using $\Sigma \approx 2\rho H_p$, the accretion rate $\dot{M}_{\rm sh}$ in our numerical units is estimated as:
\begin{align}
    \dot{M}_{\rm sh} &\sim 6.5 \times 10^{-5} \left(\frac{m_\varphi}{4}\right)\left(\frac{\psi_Q}{0.4}\right)\left(\frac{H_p/R}{0.1}\right)^3 \nonumber \\
    &\times \left(\frac{\rho}{5\times 10^{-3}\rho_0}\right)\left(\frac{R}{1.5R_\ast}\right)^{3}\left(\frac{\Omega}{0.6}\right).
\end{align}

Although uncertainties remain, $\dot{M}_{R\varphi, {\rm h}}$ (approximately $6 \times 10^{-5}$; see Figure~\ref{fig:mdot_from_angmom}) falls within the expected range of $\dot{M}_{\rm sh}$ ($3.4\times 10^{-5}$ to $7.9\times 10^{-5}$ for the given range of $\psi_Q$) at the same radius (approximately $6 \times 10^{-5}$; see Figure~\ref{fig:mdot_from_angmom}). This general consistency suggests that the spiral shocks are the primary contributor to the hydrodynamic term near the protostar. Our simulation suggests that \citet{Rafikov2016ApJ}'s theory provides a useful framework for interpreting shock-driven accretion processes.

We note that the spatial resolution of our global model is lower than those of local models used in the previous boundary layer studies \citep[for example, approximately 10 grid cells per stellar pressure scale height in][]{Belyaev2018MNRAS}. Based on the previous hydrodynamic simulations \citep[e.g.,][]{Belyaev2013ApJ}, our resolution is insufficient to accurately capture spiral shocks excited by shear-acoustic instabilities. Higher-resolution simulations are therefore necessary to derive more reliable conclusions about the shock structure.

\subsection{Decretion}\label{subsec:decretion}

\begin{figure}
    \centering
    \includegraphics[width=\columnwidth]{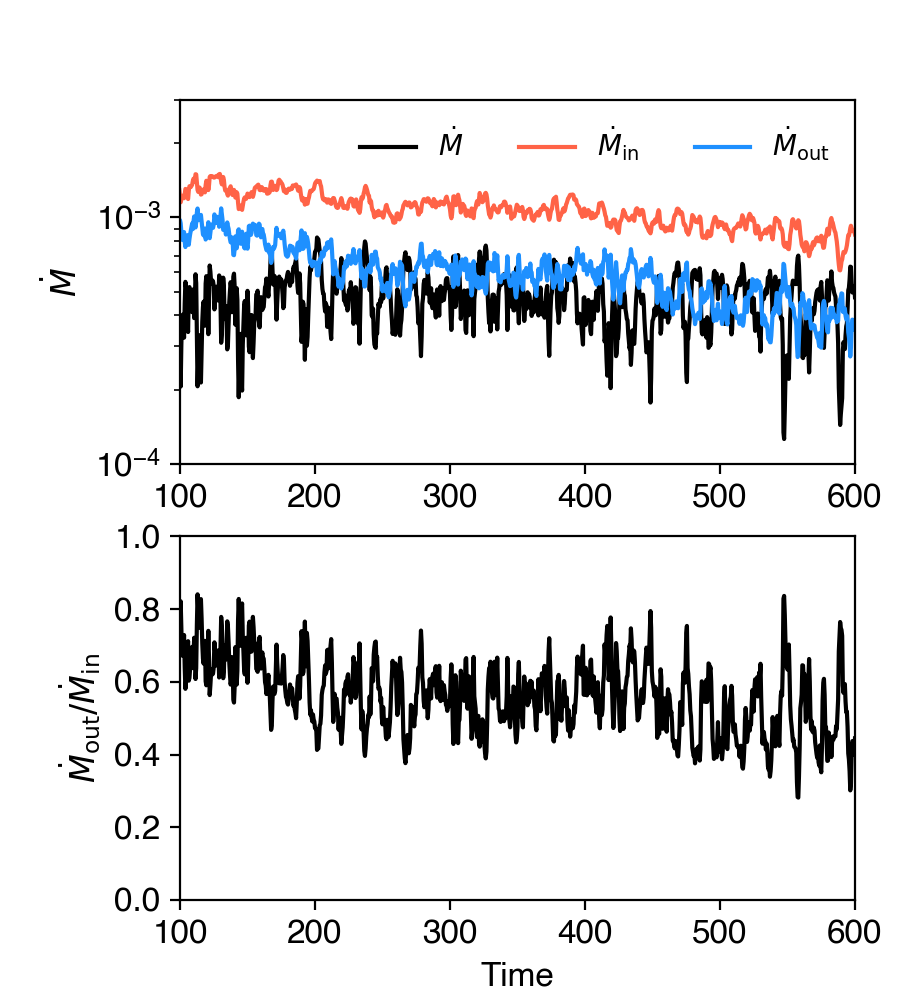}
    \caption{Top: The net accretion rate $\dot{M}$ (black), the mass inflow rate $\dot{M}_{\rm in}$ (red), and the mass outflow rate $\dot{M}_{\rm out}$ (blue). Bottom: The ratio of the outflow to the inflow rates. All data are measured at $r=2R_\ast$. Gas with a temperature below $0.1 T_0$ is considered to exclude the polar hot gas. The net accretion rate ($\dot{M}_{\rm in}-\dot{M}_{\rm out}$) is approximately $4\times 10^{-4}$, corresponding to $\sim 2.4\times 10^{-6}~{M_\odot~{\rm yr^{-1}}}$.}
    \label{fig:mdot}
\end{figure}

In Section~\ref{subsec:angmom}, we identified a decretion flow near the equatorial plane. Since decretion may play a crucial role in transporting CAIs to the outer disk, we investigate its properties here.

\citet{Takeuchi2002ApJ} found that decretion occurs near the equatorial plane when the gas density decreases sufficiently steeply with radius. Suppose that the disk is vertically isothermal, and the density and temperature scale as $\rho \propto R^{b_{\rm d}}$ and $T \propto R^{b_{\rm T}}$, where $b_{\rm d}$ and $b_{\rm T}$ are the power-law indices for the density and temperature on the midplane, respectively.
The condition for decretion is then given by:
\begin{align}
    b_{\rm d} + \frac{2}{3}b_{\rm T} < -2. \label{eq:decretion_cond}
\end{align}
In our model, $b_{\rm T}\approx n_{\rm T} = -1/2$ because we adopt the $\beta$-cooling approach. Therefore, Inequality~\eqref{eq:decretion_cond} sets the decretion condition for the density power-law index, $b_{\rm d}$.

We analyze the density profile on the equatorial plane using data at $t = 500 t_0$ (Figure~\ref{fig:1d_plots}). The density power-law index remains nearly unchanged from its initial value ($b_{\rm d} \approx n_{\rm d} = -1.875$) in the region $R / R_\ast \gtrsim 2$. In this radial range, Inequality~\eqref{eq:decretion_cond} is satisfied, and we indeed observe decretion. In contrast, the density profile decreases toward the protostar ($b_{\rm d} \gtrsim 0$) in the region $1.2 \lesssim R / R_\ast \lesssim 2$, where the condition for decretion is not met. As expected, no decretion is observed in the flat density region (Figure~\ref{fig:angflux}).

As discussed in Section~\ref{subsec:angmom}, multiple angular momentum transfer mechanisms, such as the spiral shocks, operate within a radii of a few $R_\ast$. The inner edge of the decretion zone aligns with the radius where the dominant mechanisms driving accretion change.

\citet{Jacquemin-Ide2021A&A} suggested that magnetic fields transport angular momentum from the disk surfaces toward the equatorial plane, potentially driving decretion. However, our simulation does not show a clear signature of such vertical angular momentum transport (see the bottom middle panel of Figure~\ref{fig:angflux}). Therefore, the decretion observed in our model is likely driven by radial angular momentum transport, as described in \citet{Takeuchi2002ApJ}.

To further investigate mass circulation in the radial direction, which is closely linked to decretion, we analyze the mass flow rates at $r = 2R_\ast$, shown in Figure~\ref{fig:mdot}. 
In the calculation, we only consider gas with a temperature below $0.1 T_0$ to exclude the polar hot gas.
The net accretion rate (black line in the upper panel) remains nearly constant over time. The mass inflow rate, $\dot{M}_{\rm in}$, reaches approximately $6 \times 10^{-6}~M_\odot~{\rm yr^{-1}}$, while the net accretion rate, $\dot{M}$, is approximately $2 \times 10^{-6}~M_\odot~{\rm yr^{-1}}$. 

The bottom panel of Figure~\ref{fig:mdot} shows
the ratio of outflow to inflow rates, revealing that the substantial mass (approximately 50\%) is outgoing. The outflow includes both decretion flows near the equatorial plane and outward-moving gas associated with MRI-channel flows in the accreting region. However, in this analysis, we do not distinguish between these components.

We note that the appearance and radial extent of the decretion region can depend on the properties of radiative cooling. 
\citet{Philippov2017ApJ} argued that the emergence of decretion is influenced by the vertical thermodynamic structure, a factor that was not considered in the analysis by \citet{Takeuchi2002ApJ}.
Furthermore, radiative cooling damps shock waves, which can affect the radial extent of the shock-driven accretion \citep{Dittmann2024ApJ}. This damping may alter the density profile and, consequently, modify the location and extent of the decretion region. To achieve a more comprehensive understanding of mass circulation, detailed radiative transfer calculations will be essential.

\subsection{Slow funnel accretion}\label{subsec:slow_funnel}
\begin{figure}
    \centering
    \includegraphics[width=\columnwidth]{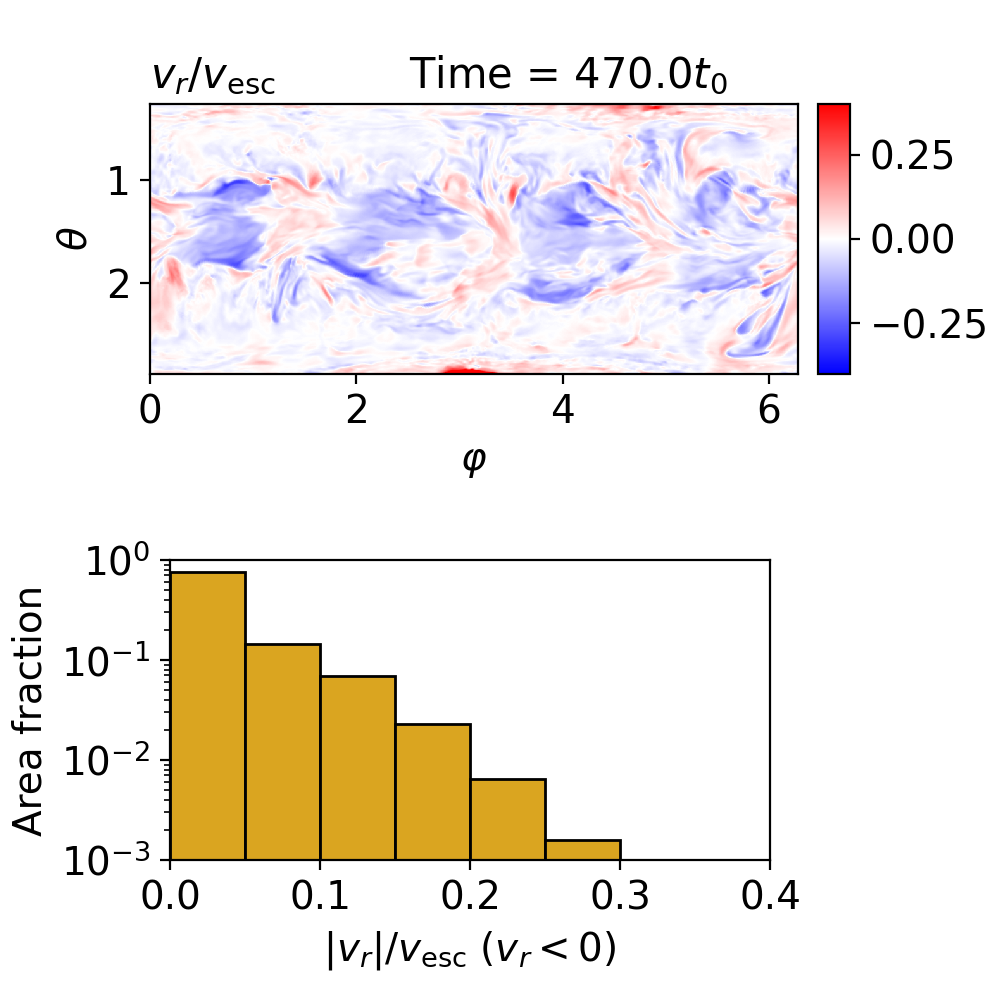}
    \caption{Top: The distribution of $v_r$ normalized by the local escape velocity $v_{\rm esc}$ on the spherical surface at $r=1.2R_\ast$. The accreting flows correspond to regions with negative values. Bottom: The area fraction histogram as a function of the accretion speed.}
    \label{fig:vr_vesc_hist}
\end{figure}

The velocity structure of accretion flows influences the properties of accretion-origin emissions, such as the hydrogen Brackett $\gamma$ line. In this section, we examine the accretion speed at high latitudes on the stellar surface, which may also be relevant to emissions from accretion shocks.

\begin{figure*}
    \centering
    \includegraphics[width=2.1\columnwidth]{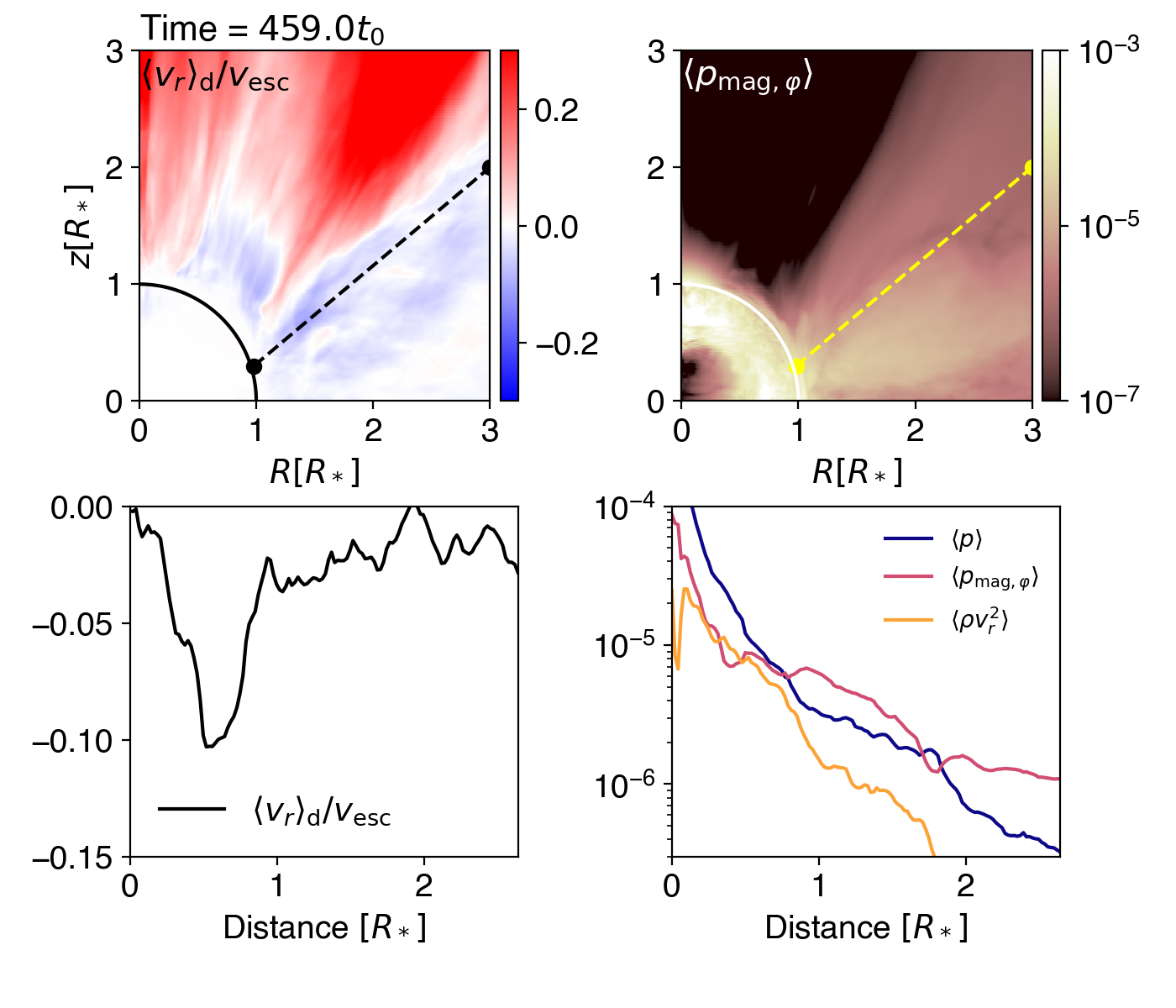}
    \caption{Top left: The density-weighted radial velocity normalized by the local escape velocity $v_{\rm esc}$. Top right: The magnetic pressure based on the toroidal fields. In both panels, the data are azimuthally averaged. The dashed lines indicate the slit for measurement, and the stellar surface is represented by the solid curves. Bottom left: $\langle v_r\rangle_{\rm d}/v_{\rm esc}$ along the slit. Bottom right: A comparison of the gas pressure $\langle p\rangle$, the magnetic pressure based on the toroidal fields $\langle p_{\rm mag,\varphi}\rangle$, and the ram pressure $\langle \rho v_r^2\rangle$ along the slit. For the 1D plots, the distance is measured from the point close to the protostar.}
    \label{fig:funnel_acc}
\end{figure*}

The top panel of Figure~\ref{fig:vr_vesc_hist} shows the $r$-$\varphi$ map of the radial velocity at $r = 1.2R_\ast$, normalized by the local escape velocity, $v_{\rm esc}(r) = \sqrt{2GM_\ast / r}$. The map reveals that the accretion speed is significantly smaller than the escape velocity, reaching at most $\sim 30$\% of it. The bottom panel displays a histogram of the area fraction on the spherical surface, illustrating that flows with speeds exceeding $30$\% of the escape velocity are extremely rare.

We compare the accretion mode in our model with other known accretion modes. In magnetospheric accretion, which is typical for CTTSs, nearly free-fall accretion flows occur along magnetic field lines connecting the inner disk to the magnetic poles of the star \citep{Hartmann2016ARA&A, Takasao2022ApJ}. Such polar accretion is absent in our model because the polar magnetic fields extend radially and do not connect to the disk (Figure~\ref{fig:magnetic_structure3d}).
Some Herbig Ae/Be stars exhibit fast accretion despite having weak magnetic fields \citep{Johnstone2014MNRAS}. \citet{Takasao2018ApJ} performed a 3D MHD simulation of accretion onto a weakly magnetized star and demonstrated that free-fall funnel accretion can occur even without a stellar magnetosphere as a result that disk winds fail to escape the system. While the magnetic geometry in their model (i.e., the absence of a stellar magnetosphere truncating the inner disk) is similar to that of the present study, our model does not exhibit free-fall funnel accretion.

We find that funnel accretion is magnetically decelerated. The top left panel of Figure~\ref{fig:funnel_acc} shows the funnel accretion structure, with a measurement slit defined along the accretion flow, as indicated by the dashed line. The bottom right panel compares the gas pressure, magnetic pressure (from the toroidal fields), and ram pressure along the slit. The analysis reveals that the ram pressure is smaller than or, at most, comparable to the magnetic pressure, consistent across different times. This result indicates that the Lorentz force from the strong toroidal magnetic fields inhibits fast accretion (see the top right panel).

The strength of the amplified magnetic fields is limited by the disk gas pressure (Figure~\ref{fig:pressure_balance}). Based on this, we infer that magnetic deceleration is more effective for rapid accretors. This inference aligns with the findings of \citet{Takasao2018ApJ}. In their model, fast accretion likely occurs because the accretion rate is significantly lower than in the present study.

\subsection{Explosive events}\label{subsec:explosive}

\begin{figure*}
    \centering
    \includegraphics[width=2\columnwidth]{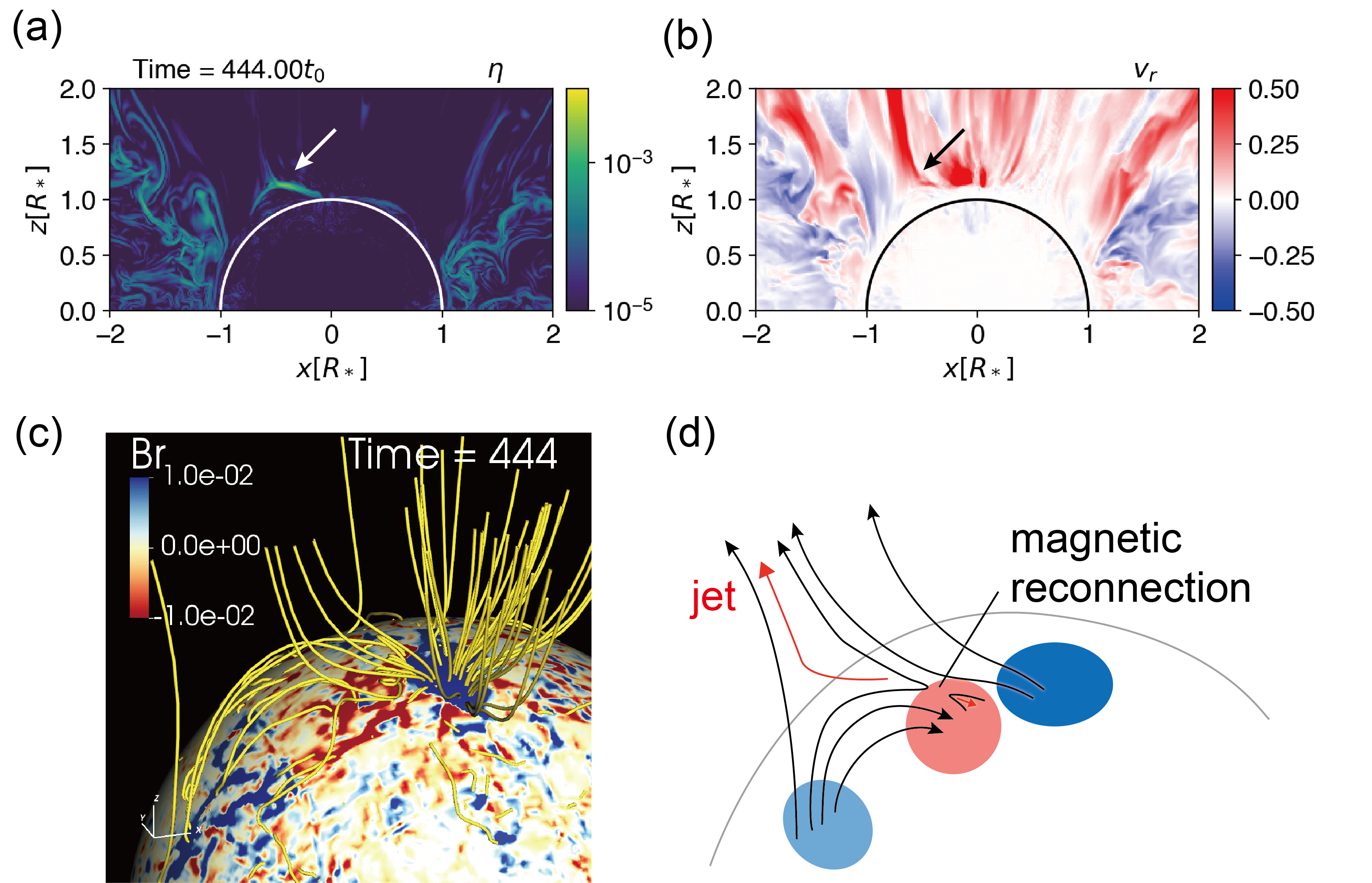}
    \caption{An example of a magnetic reconnection event between a magnetic arcade emerging from the protostellar surface and the background field. Panels~(a) and (b) display the distributions of the magnetic diffusivity $\eta$ and $v_r$ in the $xz$ plane at $y=0$, respectively. The arrows in these panels indicate the locations of the magnetic arcade driving the reconnection jet. Panel~(c) shows the 3D magnetic field structure of the magnetic arcade and the polar open fields interacting with the magnetic arcade. The stellar surface is colored by the value of $B_r$. Panel~(d) presents a schematic diagram of this reconnection event.
    }
    \label{fig:reconnection_emf}
\end{figure*}

\begin{figure*}
    \centering
    \includegraphics[width=2\columnwidth]{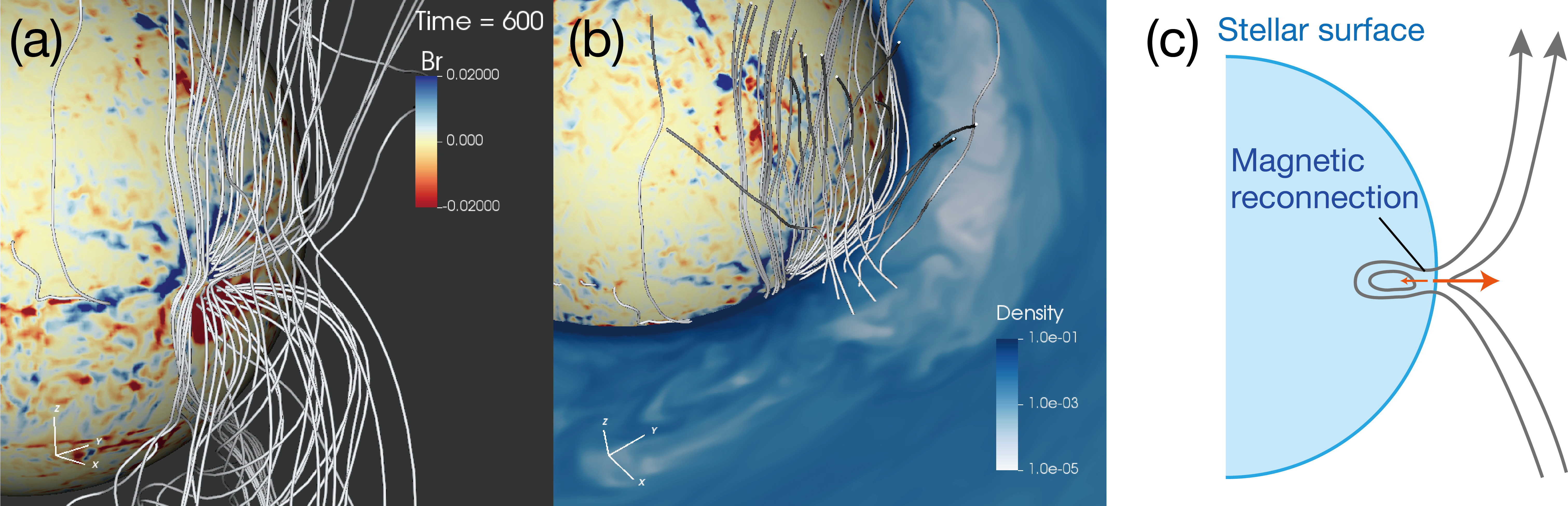}
    \caption{An example of magnetic reconnection event near the equatorial plane. (a) The stellar surface ($r=R_\ast$) is colored with the value of $B_r$. Magnetic field lines near a pair of opposite magnetic polarities are drawn. (b) The same as the panel~(a) but with the equatorial slice of the density. (c) A schematic diagram of this magnetic reconnection event.}
    \label{fig:reconnection_equator}
\end{figure*}

Figure~\ref{fig:reconnection_emf} illustrates a magnetic reconnection event near the northern pole. The resistivity map in panel~(a) reveals the formation of a strong current sheet, indicating magnetic reconnection. The $v_r$ map in panel~(b) shows a jet emanating from the reconnection site. Panel~(c) depicts the 3D magnetic field structure, where a magnetic arcade generated by stellar convection contacts the polar open magnetic fields. Panel~(d) provides a schematic illustration of this field geometry. As antiparallel field lines converge, magnetic reconnection drives jets and leads to the disappearance of the magnetic arcade, resembling solar coronal jets \citep[e.g.,][]{Shibata1994ApJ} and similar solar explosive events \citep{Takasao2013PASJ}.

Our simulation shows that magnetic reconnection frequently occurs between the polar fields and fields emerging from the protostar's interior (Figure~\ref{fig:reconnection_emf}). Stellar convection generates magnetic fields that rise into the atmosphere. However, as illustrated in Figures~\ref{fig:overview2d} and \ref{fig:magnetic_structure3d}, the protostar is enveloped by coherent magnetic fields (poloidal at the poles and toroidal at lower latitudes). These strong fields prevent the formation of stable magnetic arcades, as emerging fields undergo magnetic reconnection. Consequently, long-lived ($\gtrsim 10~t_0$) well-developed magnetic arcades, such as those seen in solar active regions, do not form (see also Figure~\ref{fig:magnetic_structure3d}).

Panels~(a) and (b) of Figure~\ref{fig:reconnection_equator} present another example of magnetic reconnection events. This event is driven by reconnection near magnetic concentrations on the protostar. These magnetic concentrations are one of four strongly magnetized regions near the equator (see Figure~\ref{fig:Brad_sphslc}), where each region exhibits a pair of opposite magnetic polarities (see also panel~(b) of Figure~\ref{fig:summary_spiral_shock}). Magnetic reconnection occurs between the opposite polarities within a magnetic concentration. Panel~(c) illustrates the magnetic geometry of this event, showing that magnetic reconnection takes place in the submerging part of the field. Similar reconnection events have been extensively studied in solar physics \citep[e.g.,][]{Isobe2007ApJ,Takasao2013PASJ}, although the spatial scale is significantly larger in this case. We find that such reconnection events become prominent after $t\approx 490~t_0$.


Magnetic reconnection near the equator drives bipolar jets. Although the reconnection jet is initially horizontally directed, the interaction with the disk gas deflects the accelerated gas toward polar directions. These reconnection-driven jets constitute a crucial unsteady component of the jets emanating from the boundary layer.
The jet speed is comparable to the Keplerian velocity at the stellar surface. It should be noted that not all eruptions are necessarily driven by magnetic reconnection. Magnetic buoyancy instability, triggered by strong toroidal fields, may also contribute to sudden eruptions \citep{Takasao2018ApJ}.



Our previous study \citep{Takasao2019ApJ} reported similar reconnection events; however, that model did not numerically resolve the stellar surface. The present model, which resolves the convective stellar surface, provides supporting evidence that protostars can exhibit explosive events powered by magnetic reconnection of accumulated poloidal fields.

Magnetic reconnection also acts to expel poloidal fields from the protostar (see panel~(c) of Figure~\ref{fig:reconnection_equator}). Indeed, the total unsigned magnetic flux measured at the stellar surface declines after such reconnection events start to frequently occur ($t\gtrsim 500t_0$). The reconnection-mediated escape of magnetic fields from a star was initially proposed by \citet{Parker1984ApJ}. 
Through repeated expulsion of open poloidal fields at different magnetic concentrations, the cavity in the inner disk gradually develops (see panel~(b) of Figure~\ref{fig:reconnection_equator}). 
This gradual growth of the inner disk cavity suggests that the stellar magnetosphere, though not yet fully developed, may be in the process of forming.
We will conduct a longer calculation to investigate the evolution of the inner disk.

We also note that magnetic reconnection occurs not only in the protostellar fields but also in the disk atmosphere. As shown in Figure~\ref{fig:overview2d}, linear structures of hot gas are ubiquitous in low-$\beta$ regions of the disk atmosphere. These linear structures trace current sheets in the turbulent disk atmosphere. In such low-$\beta$ regions, small-scale reconnection can dramatically increase the temperature, even in the disk atmosphere.

\section{Discussion}\label{sec:discussion}

\subsection{Connecting the MRI-active disk to the protostar}\label{subsec:dis_radial_struct}

\begin{figure*}
    \centering
    \includegraphics[width=1.5\columnwidth]{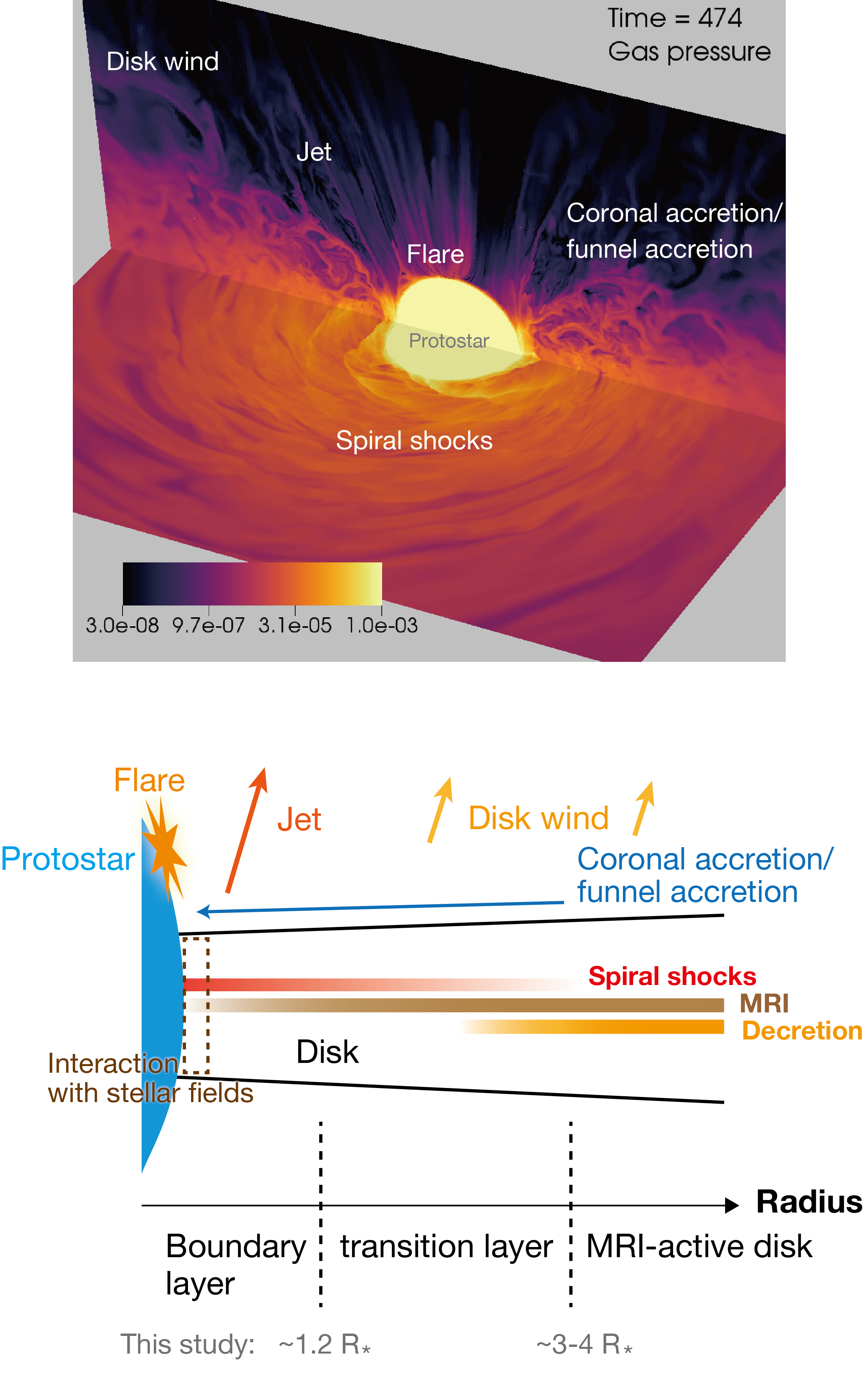}
    \caption{Top: an image of the global structure with annotations highlighting key substructures. The color represents the gas pressure in the $xy$ and $xz$ planes. The region shown corresponds to $|x| \leq 6R_\ast$, $|y| \leq 6R_\ast$, and $0 \leq z \leq 5R_\ast$. Bottom: an illustration summarizing the structure around the protostar-disk interface. The approximate radial scales identified in our model are indicated at the bottom. Note that the vertical order of the bands (spiral shocks, MRI, and decretion) holds no physical significance.}
    \label{fig:simple_summary}
\end{figure*}

Figure~\ref{fig:simple_summary} provides an overview of how the MRI-active disk connects to the protostar (see also Section~\ref{subsec:angmom}). 
The top panel presents the slices of the gas pressure distribution in 3D with annotations highlighting key substructures, while the bottom panel provides an illustration summarizing the structure around the protostar-disk interface.
Several key processes contribute to angular momentum transfer: MRI, spiral shocks, coronal accretion, jets, and disk winds. 

MRI is linearly unstable outside the boundary layer but is quenched within it. The transition between these regions is smooth, as MRI channel flows can penetrate into the boundary layer. The spiral shocks, originating from the protostellar surface, propagate outward and damp within a few stellar radii. In our model, these shocks extend beyond the boundary layer, modifying the inner MRI-active disk and forming a transitional region between the MRI-active zone and the boundary layer. The inner edge of the decretion flows on the midplane lies within the transition layer.

Above the equatorial plane, the coronal accretion flows drag magnetic fields, amplifying Maxwell stress and enhancing its role in angular momentum removal. The jets and disk winds further extract angular momentum from the disk.

\subsection{Implication for interpretation\\
of accretion-origin emissions}\label{subsec:imp_acc_emission}

The velocity structure of the accretion flows in our protostellar model differs significantly from that in magnetospheric accretion (Section~\ref{subsec:slow_funnel}). 
In magnetospheric accretion, the stellar surface is disconnected from the inner disk, as the disk is truncated by strong stellar magnetic fields. These fields channel disk material into free-fall funnel flows directed toward the magnetic poles.
In contrast, our model features a disk that extends down to the protostellar surface, forming a boundary layer. Additionally, a magnetically decelerated, slow funnel accretion flow is observed. Unlike the magnetospheric case, these funnel flows are not guided by stellar magnetic fields. Instead, they originate from the coronal accretion flows and disk winds \citep[e.g.,][]{Takasao2018ApJ}.

The difference in the accretion structures can impact interpretation of accretion-shock emissions. For instance, hydrogen Br$\gamma$ emissions are often used to estimate the accretion rates of embedded protostars \citep[e.g.,][]{Fiorellino2023ApJ}. This method relies on the empirical relation between $L_{\rm Br\gamma}$ and $L_{\rm acc}$, established from observations of CTTSs \citep{Alcala2017A&A}, where $L_{\rm Br\gamma}$ and $L_{\rm acc}$ are the Br$\gamma$ line luminosity and accretion luminosity, respectively. 

If embedded protostars exhibit magnetospheric accretion, applying this empirical relation may be valid. However, if the accretion mode differs, such a straightforward application could lead to inaccurate accretion rate estimates. In CTTSs, Br$\gamma$ emissions are thought to primarily originate from funnel accretion flows \citep{Gravity2020Natur,Gravity2023A&A}, which have temperatures of $\sim 10^4$~K, significantly higher than disk temperatures. The heating mechanism for these funnel flows, however, remains unclear \citep[see, e.g., Section~2.3 of][]{Hartmann2016ARA&A}. In our protostellar model with boundary layer accretion, numerous explosive heating events occur in the disk atmosphere (Section~\ref{subsec:explosive}), and the origin of the funnel accretion differs fundamentally from that in the magnetospheric accretion. Therefore, caution must be exercised when applying the empirical relations derived from CTTS observations, particularly when addressing the luminosity spread problem of protostars \citep{Fischer2023ASPC}.

Additionally, the speed of the funnel accretion differs between the two accretion modes. In the boundary layer accretion, the funnel accretion flow is significantly slower than the escape velocity, which may influence the accretion-shock-origin emissions (Section~\ref{subsec:slow_funnel}).

\subsection{Implication for X-ray emissions\\
in FU Ori-type stars}\label{subsec:FU-Ori}

FU Ori-type stars are rapid accretors with accretion rates of $\dot{M} \gtrsim 10^{-5}~M_\odot~{\rm yr^{-1}}$. Due to these high accretion rates, boundary layer accretion may play a significant role \citep[e.g.,][]{Labdon2021A&A}. FU Ori-type stars are also known to exhibit coronal properties distinct from those of CTTSs.

We begin by reviewing the general properties of CTTS coronae. The coronae of CTTSs are heated by both accretion and magnetic processes. The impact of accretion flows on the stellar surface generates heat in the coronae \citep{Cranmer2009ApJ}, with the temperature of the accretion shock typically reaching at most a few times 0.1~keV, based on the free-fall velocity. Magnetic heating, analogous to processes in the solar atmosphere, is likely driven by Alfv\'enic waves and magnetic reconnection \citep[e.g.,][]{Rempel2017ApJ,Shoda2021A&A}.
Non-flaring coronae of CTTSs generally exhibit temperatures of a few keV \citep[e.g.,][]{Imanishi2001ApJ,Telleschi2007A&A}. During stellar flares powered by magnetic reconnection, hot plasmas can form with temperatures reaching $\sim 10$~keV \citep[e.g.,][]{Getman2008ApJ}.

Observations of FU Ori-type stars indicate that their coronae are typically hotter and more luminous than those of CTTSs, yet they exhibit fewer strong flares.
\citet{Skinner2009ApJ} reported that an FU Ori-type star V1735 Cyg exhibited high-temperature plasma ($T > 5~{\rm keV}$) without strong flares. Since this temperature is much higher than the typical value of $\sim 0.1~{\rm keV}$ for accretion shocks, the origin of the emission must be magnetic. \citet{Kuhn2019ApJ} further showed that such quiescent high-temperature coronae are common among FU Ori-type stars. A notable feature of these stars is their significantly higher coronal temperatures compared to non-flaring CTTS coronae, suggesting differences in the underlying heating mechanisms. 
\citet{Kuhn2019ApJ} also found that FU Ori-type stars tend to be more X-ray luminous ($L_{\rm X} \gtrsim 10^{30.5}~{\rm erg~s^{-1}}$) than typical non-outbursting CTTSs, even when non-detections are considered.

Based on our simulation, we propose a possible explanation for these anomalous coronal properties of FU Ori-type stars. We suggest that the hard X-rays observed in FU Ori-type stars originate from numerous small-scale flare events in two regions: (1) the interface between emerging stellar magnetic fields and disk-originated fields (e.g., Figure~\ref{fig:reconnection_emf}) and (2) the disk atmosphere near the protostar (see Figure~\ref{fig:overview2d}). Emerging magnetic fields are likely to reconnect rapidly with surrounding toroidal or polar fossil fields before forming large magnetic arcades, which may explain the relative scarcity of strong flares compared to CTTSs. If such reconnection events occur frequently across the protostar, the cumulative effect of these small explosions could produce the quasi-steady hard X-rays observed by distant observers.
This scenario provides a plausible explanation for the formation of the very hot and luminous, yet relatively flare-less, coronae characteristic of FU Ori-type stars.

\subsection{Implication for formation and transport of CAIs}\label{subsec:CAI}

The origin of CAIs has long been a mystery. A clue lies in the presence of the short-lived radionuclide $^{10}$Be in some CAIs, which is thought to form through cosmic ray irradiation. Based on Li--Be--B isotope systematics on CAIs from carbonaceous chondrites, \citet{Fukuda2019ApJ} proposed that the precursor gas of CAIs was irradiated by a strong cosmic ray flux from the young Sun \citep[see also][]{Jacquet2019A&A}. Given that young low-mass stars are typically magnetically active \citep[e.g.][]{Getman2008ApJ}, it is plausible that the inner disk gas was exposed to such cosmic rays. However, a critical question remains: how was this irradiated gas transported outward?

\citet{Shu2001ApJ} suggested that CAIs could be formed and transported via the ``X-wind," an outflow driven by a rotating stellar magnetosphere. However, numerous issues with the assumptions and predictions of the X-wind model have been raised based on meteoritic data \citep[e.g.][]{Desch2004ApJ}.

Our simulation, despite its prescribed disk temperature profile, offers a potential alternative scenario. It suggests that cosmic-ray-irradiated gas could be transported outward by decretion flows, allowing CAIs to form in regions where the temperatures are suitable for their formation. In our model, the inner edge of the decretion region lies at a radius of a few $R_\ast$, a location where gas is likely to be irradiated by cosmic rays from the young Sun. Furthermore, the protostar in our model frequently produces magnetic reconnection events in both the protostellar and disk atmospheres (Section~\ref{subsec:explosive}), which could inject abundant cosmic rays into the inner disk. Additionally, the substantial outgoing mass flux observed in our model (Figure~\ref{fig:mdot}) may account for the mass required to produce the observed total amount of CAIs. Detailed investigations of this scenario will be our future work.

\citet{Yang2012M&PS} proposed a different scenario of a decretion disk based on a 1D viscous disk model. They studied the CAI formation and transport during the formation and growth of a protoplanetary disk in a collapsing molecular cloud. Their model is based on two key assumptions. The first assumption (assumption~(1)) is that the molecular cloud undergoes a solid body rotation, with the angular speed determined from core-scale ($\sim$0.1~pc) observations \citep[e.g.][]{Goodman1993ApJ}. The solid body rotation results in a low angular momentum in the molecular cloud core. Consequently, a small and dense disk initially forms ($\sim 1~$au when the protostellar mass is $\sim 0.4M_\odot$), where the temperature is sufficiently high for CAI formation ($\gtrsim 1,400$~K). The second assumption (assumption~(2)) is that the angular momentum in the growing disk is transported solely through viscosity. As a result of angular momentum transport due to viscosity, the disk expands radially \citep{Lynden-Bell1974MNRAS}. Namely, accretion occurs in the inner disk, while decretion occurs in the outer disk. The vertical structure of the accretion and decretion flows is ignored. This spreading disk is expected to transport CAIs from the vicinity of the proto-Sun to an outer radius. As the transition radius between decretion and accretion moves outward with time, this scenario requires a hot and compact disk in the very early phase of star formation.

However, this scenario has been challenged by observations and simulations. Observations of cloud cores ($\lesssim$0.1~pc) have found that the specific angular momentum exhibits a weaker radial dependence than that predicted by solid-body rotation \citep{Pineda2019ApJ,Gaudel2020A&A}, contradicting assumption~(1). These observations suggest that models assuming solid-body rotation underestimate the specific angular momentum of a cloud core. Indeed, Class~0/I protostars are typically observed to host disks with sizes exceeding a few tens of au \citep{Maury2019A&A,Tobin2020ApJ}, which is significantly larger than the disk predicted by \citet{Yang2012M&PS} at a comparable evolutionary stage. Furthermore, \citet{Misugi2024ApJ} conducted 3D MHD simulations and demonstrated that turbulent magnetized clouds naturally produce a weaker radial dependence of the specific angular momentum. Their simulations also showed that cloud cores retain sufficient angular momentum to form disks exceeding 10~au in size (see their Figure~16).

Furthermore, assumption~(2) may not hold for hot disks due to the significant role of magnetic fields. When the disk temperature exceeds $\sim 10^3$~K, thermal ionization of potassium (K) ensures that the ideal MHD condition is met. In such cases, the disk acquires a significant amount of magnetic fields and undergoes strong magnetic braking \citep[e.g.,][]{Mellon2008ApJ}. Magnetic fields extract angular momentum vertically from the disk, thereby preventing the disk from expanding radially. The formation of Keplerian disks can also be prevented, a phenomenon known as the magnetic braking catastrophe \citep[e.g.,][]{Li2014prpl}.

Our decretion disk significantly differs from the 1D viscous disk model. In our model, decretion occurs even in the inner disk where the net mass flow is inward. The decretion and accretion coexist, suggesting that radial transport may be possible even after the development of ten-au-scale disks. The development of such disks indicates that they have avoided significant magnetic braking, implying that the plasma $\beta$ of the disk vertical fields is considerably larger than unity. Our disk has an initial plasma $\beta$ of $10^3$. Therefore, although our disk is assumed to be well-coupled with magnetic fields, it maintains Keplerian rotation outside the boundary layer. Based on our model, we consider that the formation and radial transport of CAIs are not limited to the very early phase of star formation where the disk size is 1 au or less.

\subsection{Model limitations}\label{sec:model_limitations}

We first discuss potential impacts of stellar rotation. In our model, the protostar is initially non-rotating. For rotating protostars, the strength of spiral shocks is expected to diminish due to a smaller velocity difference between the protostellar surface and the disk. The pressure disturbance resulting from the collision between the rotating disk gas and the magnetically modulated stellar surface is proportional to the ram pressure, $\delta p \propto \rho (v_{\rm K} - v_\ast)^2$, where $v_\ast$ is the rotation speed of the protostellar surface. If this relation holds, spiral shocks will remain important for angular momentum transfer, particularly in slowly rotating protostars. For rapidly spinning protostars, jets originating from the boundary layer may dominate the accretion-driving mechanisms. Additionally, spinning protostars are likely to power polar jets \citep[e.g.,][]{Machida2008ApJ}, which are absent in our model. Stellar rotation could also enhance dynamo activity \citep{Wright2011ApJ}.

Our model employs a simplified cooling function (Section~\ref{subsec:basic_eq}), which imposes artificial constraints on the specific entropy of the accreting gas. The heat injection rate is a critical factor influencing protostellar evolution \citep{Palla1992ApJ,Hosokawa2011ApJ,Baraffe2012ApJ,Kunitomo2017A&A}, and the prescribed thermal structure in our model does not account for the effects of radial energy flux, which can significantly alter the inner disk structure \citep{Popham1993ApJ}. Incorporating radiative transfer is essential to improve the realism of the model. Furthermore, as discussed in Section~\ref{subsec:spiral_shocks}, the cooling properties are likely to influence the behavior of the spiral shocks.

The formation of magnetic concentrations is likely to occur in reality. However, we must note that the development of the $m_{\varphi}=4$ mode in our model is probably an artifact. We attribute this to the influence of the Cartesian grid on the protostellar convection pattern, as observed in previous studies \citep[e.g.,][]{Ott2012PhRvD}. 

In the convective layer, low-entropy flows typically correspond to downflows, while high-entropy flows correspond to upflows. Numerical dissipation, which increases entropy, is less significant for flows aligned with the grid directions but more pronounced for flows at angles to the grid. Consequently, flows parallel to the grid become less buoyant and tend to form downflows. Our simulation shows stronger downflows along the grid directions, leading to the formation of the $m_{\varphi}=4$ large-scale magnetic concentrations (see the $v_r$ map in Figure~\ref{fig:spiral_shocks2d}). Although the mechanism by which protostellar magnetic fields excite the spiral shocks appears plausible, our simulation cannot reliably predict the number of spiral shocks. This limitation highlights the influence of grid geometry on our results.

Numerical resolution is another critical factor in simulations of turbulent plasmas. In our model, the smallest grid size is approximately $10^{-2}R_\ast$, which is insufficient for resolving the stellar dynamo with numerical convergence. While accurately resolving the dynamo is beyond the scope of this study, uncertainties in the generation of the stellar magnetic fields must be acknowledged. Additionally, the minimum grid size in our model is comparable to the pressure scale height at the stellar surface. \citet{Belyaev2012ApJ} demonstrated that resolving the sonic instability in the boundary layer requires a grid size much smaller than the scale height.

\section{Summary}\label{sec:summary}

We conducted a global 3D MHD simulation to investigate the interaction between a magnetized convective protostar and an MRI-active disk. The global nature of our model enabled us to study the critical role of vertical angular momentum transport, which could not be explored using vertically unstratified local models. By resolving stellar convection, we uncovered its significant contributions to angular momentum transport and the generation of stellar magnetic fields.


Below, we summarize our main findings:
\begin{enumerate}
    \item {\bf Radial structure of angular momentum transfer mechanisms:} Our study revealed that angular momentum transport mechanisms vary significantly from the outer disk to the protostellar surface (Sections~\ref{subsec:angmom} and \ref{subsec:angmom2}). The boundary layer forms within a radius of $\sim 1.2R_\ast$. Beyond $\sim 3{\text -}4R_\ast$, MRI primarily drives accretion. Within that radius, however, multiple mechanisms contribute to angular momentum transport: MRI, spiral shocks, coronal accretion, jets, and disk winds (Figure~\ref{fig:summary_spiral_shock}). The interplay of these mechanisms results in three distinct disk structures: (1) the MRI-active disk, (2) the transition layer between the MRI-active disk and boundary layer, and (3) the boundary layer (Figure~\ref{fig:simple_summary} in Section~\ref{subsec:dis_radial_struct}). These findings underscore the critical importance of global MHD models in understanding angular momentum transfer.
    \item {\bf Discovery of magnetically excited spiral shocks:} The protostar exhibits magnetic concentrations, analogous to starspots, generated by convection. Spiral shocks emanate from these starspots near the equatorial plane, driving accretion (Section~\ref{subsec:spiral_shocks}). These shocks are triggered by collision between the rotating disk gas and magnetic bumps on the protostellar surface, highlighting the critical role of stellar magnetism in shaping the inner disk structure.
    \item {\bf Magnetism of convective protostar:} The protostar in our model is strongly magnetized through a combination of poloidal field accumulation around the poles, wrapping by toroidal fields, and convective dynamo activity (Section~\ref{sec:overview_mag}). The polar fields are open to outer space and do not connect to the disk. The magnetic pressure of the toroidal fields is comparable to the gas pressure at the disk surface. The protostar forms multiple magnetic concentrations, resembling starspots, driven by convection (Section~\ref{subsec:spiral_shocks}). However, in our model, the magnetic fields of the major concentrations are primarily disk-field-origin rather than stellar-dynamo-origin. Although the protostar generates magnetic fields through a convective dynamo, it does not develop long-lived, well-structured magnetic arcades because the emerging fields quickly reconnect with the ambient fields (Section~\ref{subsec:explosive}).
    \item {\bf Explosive events powered by magnetic reconnection:} Numerous explosive events are driven by magnetic reconnection (Section~\ref{subsec:explosive}). Dynamo-generated magnetic arcades produce hot jets in the polar regions, analogous to solar coronal X-ray jets. Reconnection also occurs in magnetic concentrations near the protostellar equator, generating bipolar jets that act as unsteady components of the jets emanating from the boundary layer. These findings are consistent with our previous model \citep{Takasao2019ApJ}, where the stellar surface was treated as the inner boundary. Additionally, we observe numerous small-scale reconnection events in low-$\beta$ regions of the disk atmosphere, which locally increase the temperature. These explosive events may provide valuable insights into the origins of hot emissions from accreting young stars, such as X-rays, and could help explain the anomalous X-ray properties observed in FU Ori-type stars (Section~\ref{subsec:FU-Ori}).
    \item {\bf Emergence of decretion region:} Our disk exhibits decretion on the midplane beyond $R \gtrsim 2{\text -}3R_\ast$ (Section~\ref{subsec:decretion}). The inner edge of the decretion zone aligns with the radius where the driving mechanisms of accretion transition. The coexistence of accretion and decretion layers (i.e. meridional circulation) results in strong mass circulation within the disk. Decretion flows may play a crucial role in the radial transport of precursor gas for CAIs (Section~\ref{subsec:CAI}). Meteoritic data analyses suggest that precursor gas of CAIs was irradiated by cosmic rays produced by flares of the young Sun. Our protostellar model may support this scenario.
    \item {\bf Absence of nearly free-fall accretion flows:} Many previous models of magnetized accretion disks show nearly free-fall accretion flows above the disk surfaces, which are expected to collide with the protostar if there are no obstacles. However, in our model, such fast accretion flows do not occur (Section~\ref{subsec:slow_funnel}). We found that strong toroidal fields wrapping the protostar act to prevent rapid accretion. This finding may have implications for interpreting observations of accretion-origin emissions, such as the hydrogen Brackett $\gamma$ line (Section~\ref{subsec:imp_acc_emission}).
\end{enumerate}

We thank Drs. Kohei Fukuda, Sota Arakawa, Yoshiaki Misugi, and Yusuke Tsukamoto for useful comments.
This research could never be accomplished without the support by Grants-in-Aid for Scientific Research (ST: JP22K14074; ST, KI and KT: JP21H04487, JP22KK0043; TH: JP19KK0353, JP22H00149) from the Japan Society for the Promotion of Science. 
TH was financially supported by ISHIZUE 2024 of Kyoto University.
Numerical computations were carried out on Cray XC50 at Center for Computational Astrophysics, National Astronomical Observatory of Japan.
This work was partly achieved through the use of large-scale computer systems at the Cybermedia Center, Osaka University.
This work was also supported by MEXT as a Program for Promoting Researches on the Supercomputer Fugaku ``Structure and Evolution of the Universe Unraveled by Fusion of Simulation and AI" (Grant Number JPMXP1020230406).

\appendix

\section{Details of cooling term}\label{app:cooling}

We describe how the reference temperature $T_{\rm ref}$ and cooling timescale $\tau_{\rm cool}$ are calculated in our $\beta$-cooling model.

To construct the reference temperature $T_{\rm ref}$, we aim to smoothly connect $T_\ast(r)$ and $T_{\rm d}(R)$. However, predicting the boundary between the rotating disk gas and the pressure-supported stellar gas in advance is challenging. Since the star-disk boundary evolves during the simulation, the initial temperature profile of the system is unsuitable as a reference temperature. Therefore, we design a functional form for $T_{\rm ref}$ that adapts to the evolving system.

We distinguish between the two gas regimes using a dynamical criterion: the ratio of the centrifugal force to the gravitational force, $X_c = (v_\varphi / v_{\rm K}(R))^2$. When the gas experiences a significant centrifugal force, we assume that the reference temperature follows $T_{\rm d}(R)$; otherwise, it follows $T_\ast(r)$. In this study, we assume that protostellar gas satisfies $X_c < 0.1 \equiv X_{c,0}$. Based on these requirements, $T_{\rm ref}$ is constructed using a switching function $f_{\rm rot}$:
\begin{align}
    T_{\rm ref} = T_{\rm ref}(v_\varphi, \bm{r}) = f_{\rm rot}T_{\rm d}(R) + (1 - f_{\rm rot})T_\ast(r),
\end{align}
where
\begin{align}
    f_{\rm rot} = f_{\rm rot}(R, v_\varphi) = \frac{1}{2} \left[1 + \tanh{\left( \frac{X_c - X_{c,0}}{\Delta X_c} \right)} \right].
\end{align}
We adopt $\Delta X_c = 0.02$. The switch $f_{\rm rot}$ approaches unity when $X_c > X_{c,0}$. While the choice of $X_{c,0}$ may influence the resulting boundary layer structure, a detailed investigation of this effect is left for future studies.

We design the functional form of the cooling timescale $\tau_{\rm cool}$ to satisfy the following requirements: (1) the cooling term is turned off when $T < T_{\rm ref}$, (2) the cooling term operates only near and outside the protostellar surface, and (3) the cooling term does not cause numerical instabilities in low-density regions. For simplicity, we assume that $\tau_{\rm cool}$ scales with the Keplerian timescale $t_{\rm K}(r) = 2\pi / \Omega_{\rm K}(r)$. Using a nondimensional coefficient $f_{\rm cool}$ and a switching function $F_{\rm swtc}$, we define the cooling timescale as:
\begin{align}
\tau_{\rm cool} = f_{\rm cool} t_{\rm K}(r) F_{\rm swtc}^{-1},
\end{align}
where we set $f_{\rm cool} = 0.2$. 

To meet the above requirements, we introduce three switching functions ($F_T$, $F_p$, and $F_\rho$) and express $F_{\rm swtc}$ as:
\begin{align}
F_{\rm swtc} = F_T F_p F_\rho.
\end{align}

To ensure that the cooling term $\Lambda$ only takes non-positive values (i.e., the requirement (1): no radiative heating occurs), we define the switching function $F_T$ as:
\begin{align}
    F_T = \begin{dcases}
        1 & (\text{if}~T > T_{\rm ref}), \\
        0 & (\text{otherwise}).
    \end{dcases}
\end{align}
This condition prevents radiative heating from weakening or stopping convective downflows at the stellar surface.

Regarding requirement (2), we first use the gas pressure to characterize the stellar surface. Using the initial stellar surface pressure $p_{\ast,0}$ and a nondimensional parameter $f_p (>1)$, we require that the cooling term operates only when the gas pressure $p$ satisfies $p > f_p p_{\ast,0} \equiv p_{\rm cool}$. In this study, we set $f_p = 5$.

Additionally, we ensure that the protostellar structure relaxes to a hydrostatic solution for the protostellar temperature profile when accretion stops. The reference pressure profile, $p_{\rm ref}$, is taken from the hydrostatic stellar envelope solution for $T_\ast(r)$. For simplicity, the solution is replaced with the isothermal solution for $T = T_{\rm ps}$ for $r \ge R_\ast$.

To satisfy these requirements, we define the functional form of $F_p$ near the stellar surface as follows. If the gas rotation is significantly sub-Keplerian, we set:
\begin{align}
    F_p &= \frac{1}{2}\left[1 + \tanh{\left( \frac{p/p_{\rm ref} - c_{p,1}}{\Delta_{p,1}} \right)} \right] \nonumber \\
    &\times \frac{1}{2}\left[1 - \tanh{\left( \frac{p/p_{\rm cool} - c_{p,2}}{\Delta_{p,2}} \right)} \right], \label{eq:F_p}
\end{align}
where $c_{p,1} = 1.06$, $\Delta_{p,1} = 5 \times 10^{-3}$, $c_{p,2} = 1$, and $\Delta_{p,2} = 0.1$.
If the gas is rotating rapidly, it is considered disk gas, and we set $F_p = 1$. The threshold for gas rotation is $f_{\rm rot} = 0.1$.
For $p = p_{\rm ref}$ and $p = 1.065 p_{\rm ref}$, the switch (Equation~\ref{eq:F_p}) takes values of $F_p = \mathcal{O}(10^{-5})$ and $0.90$, respectively. This indicates that $F_p$ is activated only when $p > p_{\rm ref}$.

Regarding requirement (3), the cooling term can cause numerical instabilities in very low-density regions by reducing the plasma $\beta$. To prevent such issues, we introduce the switching function $F_\rho$, which deactivates the cooling term when the local density is much smaller than the initial disk midplane density (Equation~\ref{eq:rho_mid}):
\begin{align}
    F_\rho(\rho, \bm{r}) &= \frac{1}{2}\left[1 + \tanh{\left( \frac{\rho / \rho_{\rm cool}(R) - 1}{\Delta_\rho} \right)} \right], \\
    \rho_{\rm cool}(R) &= f_{\rho,c} \rho_{\ast,0} \left(\frac{R}{R_\ast}\right)^{n_{\rm d}},
\end{align}
where $f_{\rho,c} = 10^{-5}$ and $\Delta_\rho = 0.1$ in this study.

The final switching function $F_{\rm swtc}$ is then defined as:
\begin{align}
    F_{\rm swtc} &= F_{\rm swtc}(\rho, T, v_\varphi, \bm{r}) \nonumber \\
    &= F_T(T, v_\varphi, \bm{r}) F_p(\rho, T, r) F_\rho(\rho, \bm{r}),
\end{align}
with the dependencies of the individual switches explicitly shown.

\section{Artificial magnetic diffusivity}\label{app:artificial_diffusivity}

\begin{figure*}
    \centering
    \includegraphics[width=2\columnwidth]{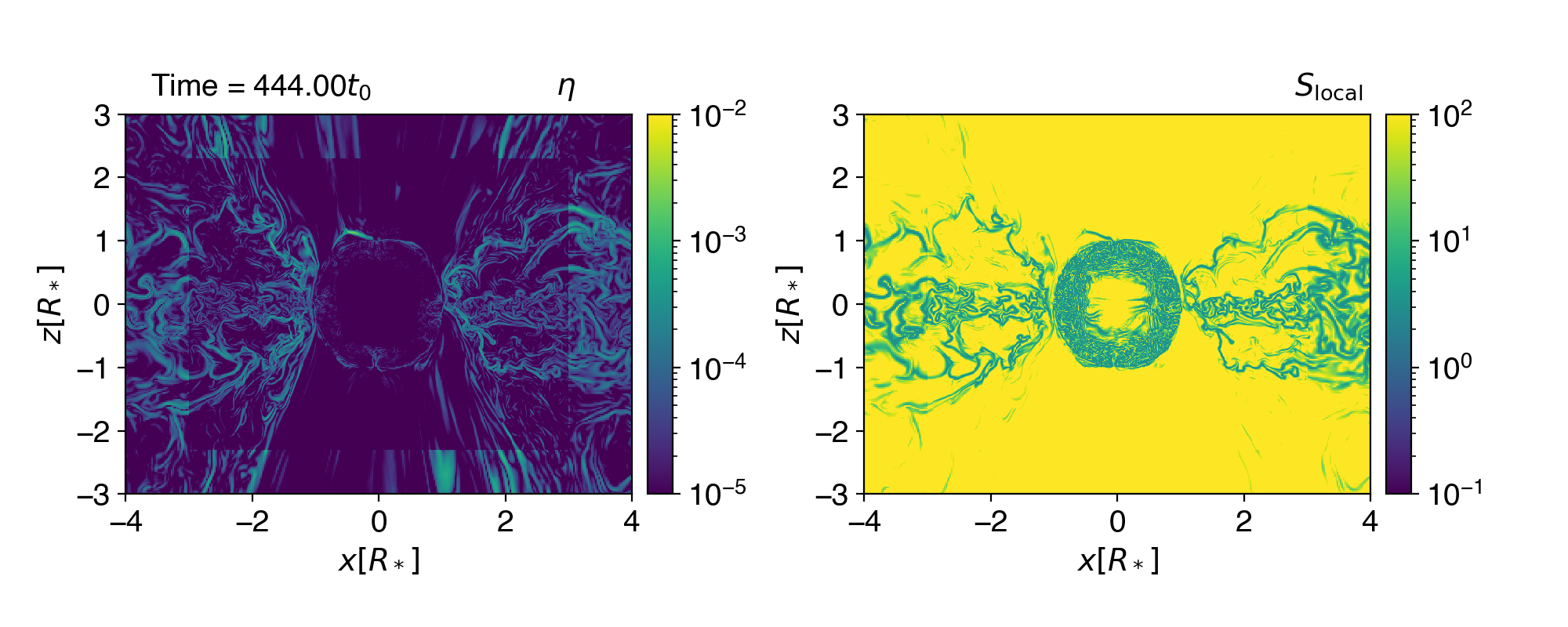}
    \caption{The artificial magnetic diffusivity $\eta$ (left) and local Lundquist number $S_{\rm local}$ (right) on the $xz$ plane ($y=0$).}
    \label{fig:diffusivity}
\end{figure*}

Figure~\ref{fig:diffusivity} illustrates the behavior of our artificial magnetic diffusivity in the simulation. The figure shows that $\eta$ takes large values only within narrow electric current sheets. The local Lundquist number, $S_{\rm local}$, reaches its minimum value of $c_\eta^{-1}$ in these regions. Outside the current sheets, $S_{\rm local}$ remains much greater than unity, confirming that the artificial diffusivity is negligible outside the narrow current sheets. Consequently, the diffusivity does not affect the dynamics in regions beyond the sharp electric current sheets.

The level boundaries of the grids are located at the outer edges of the fine mesh blocks in the range $-3 \lesssim x/R_\ast \lesssim 3$ and $-2.2 \lesssim z/R_\ast \lesssim 2.2$. Despite these level boundaries, the local Lundquist number does not show any discontinuities, demonstrating that our formulation functions effectively with mesh refinement.

\bibliography{references}{}
\bibliographystyle{aasjournal}



\end{document}